# ASSESSING THE COLLISION NATURE OF CORONAL MASS EJECTIONS IN THE INNER HELIOSPHERE

WAGEESH MISHRA
*CAS Key Laboratory of Geospace Environment, Department of Geophysics and Planetary Sciences, University of Science and Technology of China, Hefei 230026, China*

YUMING WANG
*CAS Key Laboratory of Geospace Environment, Department of Geophysics and Planetary Sciences, University of Science and Technology of China, Hefei 230026, China*

NANDITA SRIVASTAVA[*]
*Udaipur Solar Observatory, Physical Research Laboratory, Badi Road, Udaipur 313001, India*

CHENGLONG SHEN
*CAS Key Laboratory of Geospace Environment, Department of Geophysics and Planetary Sciences, University of Science and Technology of China, Hefei 230026, China*

## ABSTRACT

There have been few attempts in the past to understand the collision of individual cases of interacting Coronal Mass Ejections (CMEs). We selected 8 cases of interacting CMEs and estimated their propagation and expansion speeds, direction of impact and masses exploiting coronagraphic and heliospheric imaging observations. Using these estimates with ignoring the errors therein, we find that the nature of collision is perfectly inelastic for 2 cases (e.g., 2012 March and November), inelastic for 2 cases (e.g., 2012 June and 2011 August), elastic for 1 case (e.g., 2013 October) and super-elastic for 3 cases (e.g., 2011 February, 2010 May and 2012 September). Admitting large uncertainties in the estimated directions, angular widths and pre-collision speeds; the probability of perfectly inelastic collision for 2012 March and November cases diverge from 98%-60% and 100%-40%, respectively, reserving some probability for other nature of collision. Similarly, the probability of inelastic collision diverge from 95%-50% for 2012 June case, 85%-50% for 2011 August case, and 75%-15% for 2013 October case. We note that probability of super-elastic collision for 2011 February, 2010 May and 2012 September CMEs diverge from 90%-75%, 60%-45% and 90%-50%, respectively. Although the sample size is small, we find a good dependence of nature of collision on CMEs parameters. The crucial pre-collision parameters of the CMEs responsible for increasing the probability of super-elastic collision, in descending order of priority, are their lower approaching speed, higher expansion speed of the following CME over the preceding one, and longer duration of collision phase.

Corresponding author: Wageesh Mishra
wageesh@ustc.edu.cn, ymwang@ustc.edu.cn, nandita@prl.res.in, clshen@ustc.edu.cn

[*] Centre for Excellence in Space Sciences, India, http://www.cessi.in

1. INTRODUCTION

Coronal Mass Ejections (CMEs) are episodic expulsions of magnetized plasma from the Sun into the heliosphere and were discovered in 1970s (Hansen et al. 1971; Tousey 1973). They are the drivers of major space weather events that pose danger to space and ground based technology. The different parts (i.e., sheath, shock and cloud) of a CME have different effects on Earth's magnetosphere (Tsurutani et al. 1988; Gonzalez et al. 1994; Echer et al. 2008). In the last few decades, a significant progress has been made in understanding CMEs, including their morphological and kinematic evolution in the heliosphere, using observations from a series of imaging instruments located in space and at ground combined with modeling efforts (Lindsay et al. 1999; St. Cyr et al. 2000; Zhao et al. 2002; Xie et al. 2004; Yashiro et al. 2004; Schwenn et al. 2005; Schwenn 2006; Vršnak et al. 2010; Chen 2011; Webb & Howard 2012). It is yet impossible to forecast when a CME will be launched from the Sun and difficult to forecast accurately its arrival time at a particular location in the heliosphere. Thus, accurate space weather forecasting remains a difficult task (Tucker-Hood et al. 2015; Hess & Zhang 2015; Möstl et al. 2015). Prior to having a facility of continuous imaging of the vast distance between the Sun and Earth using twin *Solar TErrestrial RElations Observatory (STEREO)* spacecraft (Kaiser et al. 2008), CMEs were termed as ICMEs (Interplanetary CMEs) when detected away from the Sun using in situ instruments, i.e., at 1 AU. Various studies have identified the signatures of CMEs in in situ observations based on the magnetic field, velocity, temperature, density, plasma composition, plasma wave and suprathermal particles, etc. (Borrini et al. 1982; Klein & Burlaga 1982; Gosling et al. 1987; Gloeckler et al. 1999; Lepri et al. 2001; Cane & Richardson 2003; Zurbuchen & Richardson 2006; Richardson & Cane 2010). However, if CMEs interact or collide with any other large-scale solar wind structures, their in situ signatures are modified and found to be different than the signatures of a typical individual CME. The interacting CMEs structures are classified as compound streams or multiple ejecta (Burlaga et al. 1987, 2002; Wang et al. 2002).

The possibility of CME-CME interaction was pointed out by Intriligator (1976) using the in situ solar wind observation from *Pioneer* 9 and 10 spacecraft. A typical individual CME passes over Earth in around 20 hr while some structures take around several days and are possibly formed out of multiple CMEs (Marubashi & Lepping 2007; Dasso et al. 2009). The resulted complex structure from multiple CMEs may deposit its energy into the Earth's magnetosphere for long duration and lead to intense geomagnetic storms (Wang et al. 2003; Farrugia & Berdichevsky 2004; Farrugia et al. 2006; Lugaz & Farrugia 2014). CME-CME interaction has been studied for more than a decade since its first observational evidence provided by Gopalswamy et al. (2001) using the Large Angle and Spectrometric Coronagraph (LASCO; Brueckner et al. 1995) on board the *SOlar and Heliospheric Observatory (SOHO)*. The magnetohydrodynamics (MHD) numerical simulations have attempted to address the physical mechanism in CME-CME interaction, CME-CME driven shock interactions and their consequences (Vandas et al. 1997; Vandas & Odstrcil 2004; Gonzalez-Esparza et al. 2004; Lugaz et al. 2005; Wang et al. 2005; Xiong et al. 2006, 2007, 2009; Lugaz et al. 2013; Shen et al. 2013, 2014; Niembro et al. 2015; Shen et al. 2016). Realizing the importance of studying the CME-CME interaction and on availability of wide angle imaging observations of Heliospheric Imagers (HIs) with Coronagraphs (CORs) on board SECCHI/*STEREO*, almost a dozen cases of interacting CMEs have been reported in the literature in the last 5 years (e.g., Harrison et al. 2012; Liu et al. 2012; Lugaz et al. 2012; Möstl et al. 2012; Martínez Oliveros et al. 2012; Shen et al. 2012; Temmer et al. 2012; Webb et al. 2013; Mishra & Srivastava 2014; Ding et al. 2014; Lugaz & Farrugia 2014; Mishra et al. 2015a; Colaninno & Vourlidas 2015). These studies based on simulations and observations have discussed the evolution of CME-driven shock, resulted structure, nature of CME-CME collision, particle acceleration as well as geoeffectiveness. The precise information about the nature of CME-CME collision may help in determining the change between their pre- and post-collision dynamics. The use of post-collision dynamics is expected to give more accurate arrival times of CMEs at the Earth than the use of pre-collision dynamics (Mishra et al. 2015a).

Earlier studies on different candidate of interacting CMEs, using the mass and kinematics estimates from multiple viewpoint observations of *STEREO*, have posed a question as to what decides the nature of collision to vary from super-elastic (Shen et al. 2012, 2013; Colaninno & Vourlidas 2015; Shen et al. 2016) to inelastic (Temmer et al. 2012; Lugaz et al. 2012; Mishra & Srivastava 2014; Mishra et al. 2015a,b). There were several limitations in some of these studies, such as no consideration of oblique collision scenarios in three-dimension (3D), expansion speeds, and angular widths of the CMEs. These limitations and the uncertainties were only addressed in detail by Shen et al. (2012) and Mishra et al. (2016). Although, Shen et al. (2012) is a milestone on the way of our understanding the nature of CME-CME collision, their study does not attempt to constrain the conservation of momentum to remain valid for the collision scenario as in Mishra et al. (2016) admitting the involved errors in the observed characteristics of the CMEs.

The present study is the next step after Mishra et al. (2016) in understanding the nature of collision of CMEs by taking several cases of the colliding CMEs. Similar to the study of Mishra et al. (2016), we look into the role of CME characteristics, e.g., propagation direction, propagation speed, expansion speed and angular size, in deciding the collision nature of CMEs. We attempt to find the uncertainties involved while assessing the nature of collision of the CMEs by estimating the coefficient of restitution (Newton 1687) value. We determine the CMEs characteristics using *STEREO* and *SOHO* observations and assess how a reasonable uncertainty in the measured characteristics changes the probability of one type of collision nature to the other. Section 2 describes the selection of CME events, their tracking in the heliosphere using the available imaging observations and estimating their characteristics (i.e. kinematics and mass) in the pre- and post-collision phases. The analysis for the coefficient of restitution ($e$) for the selected cases and their interpretation is given in Section 3. The obtained results from all the cases are summarized in Section 4. The limitations involved in the present study is discussed in Section 5 and the conclusions are presented in Section 6.

## 2. SELECTION OF EVENTS

We first selected all the pair of CMEs launched in quick succession from almost the same source region on the Sun in *STEREO* era up to year 2013 which were identified as front-side halos or partial halos in *SOHO*/LASCO images. The CMEs of *STEREO* era were chosen as their 3D parameters could be estimated. Further, we chose only those cases where the following CME was having larger speed than the preceding CME and collision was expected at beyond couple of solar radii from the Sun. The role of magnetic forces not considered in present study are important close to the Sun, therefore the cases likely to have the collision close to the Sun are not included. Finally, we selected only those interacting cases of the CMEs which could be clearly tracked in the heliosphere by at least one heliospheric imager (HI) onboard *STEREO*. Following this, in the present study, a total of 8 cases of interacting CMEs are selected which collided with one another before reaching to Earth. Although the selected cases are still limited in number, we think that it may be extremely lengthy and difficult to select enough collision cases to perform a statistically significant study.

The selected CMEs could be tracked from the corona to the collision sites or beyond using the imaging instruments onboard *STEREO* spacecraft. The details of these selected cases of colliding CMEs are listed in Table 1. These 8 events are classified into three categories based on three criteria: (i) availability of their observations from multiple viewpoints, (ii) distance of the collision sites from the Sun, (iii) feasibility of marking the complete phase of collision duration. The duration of "collision" refers to the time interval during which an exchange of momentum between the CMEs takes place as described in our earlier studies (Mishra et al. 2015a, 2016). The estimates for observed collision duration of CMEs is not always precise. This is due to the poorly identified boundary of the collision directly from the observations. The errors in identifying the start and end of the collision lead to the errors in the measured pre- and post-collision dynamics of the CMEs. For two cases of the CMEs selected in our study, the end of the collision phase could not be identified. This causes some compromise in the accuracy of the estimated post-collision dynamics of such CMEs. The 3D kinematics of a CME can be estimated using only the single viewpoints observations of HIs (Kahler & Webb 2007; Lugaz et al. 2009; Davies et al. 2012). This is because HIs image a CME at and across large distance from the Sun where geometrical and Thomson scattering linearity break down (Howard 2011). However, earlier studies have shown that stereoscopic reconstruction methods applied on HIs observations from multiple viewpoints of *STEREO* are more accurate than single spacecraft reconstruction methods for the estimation of the kinematics of CMEs (Liu et al. 2010a; Lugaz et al. 2010; Davies et al. 2013; Mishra & Srivastava 2013; Mishra et al. 2014). Three of the selected events in our study were not well observed from both viewpoints of *STEREO*/HI and therefore we have to use single spacecraft reconstruction methods for those cases. The accuracy of the kinematics estimated using only the single viewpoint observations would be limited. We point out that the CMEs colliding at far from the Sun involves large errors in their tracking and reconstruction causing large uncertainties in their estimated kinematics (Wood et al. 2010; Liu et al. 2010b; Davies et al. 2012; Mishra et al. 2014, 2015b). The kinematics with limited accuracy would tend to reduce the accuracy of the analysis for those CMEs. Thus, accuracy of our analysis will be highest for the cases where the three criteria as aforementioned are met favorably by the CMEs, i.e., heliospheric observations from both HI-A and B are available, the collision site was not too far away from the Sun, and collision phase could be well distinguished.

Table 1 shows that all the three criteria are met favorably for the cases of 2011 February 14-15 and 2012 June 13-14 CMEs. Thus these 2 cases are with the highest accuracy in our analysis. The cases of 2010 May 23-24, 2012 March 4-5, 2012 November 9-10 and 2013 October 25 CMEs met favorably only two criteria. Therefore, the analysis for these 4 cases are considered with a moderate accuracy. The colliding CMEs of 2011 August 3-4 and 2012 September 25-28 satisfy only criteria favorably and are noted to have the lowest accuracy among the cases selected for our study.

### 2.1. *Tracking of the CMEs and Estimation of their Kinematics in COR and HI Field of View*

In this section, we track the CMEs in the heliosphere using the observations of *STEREO* coronragraphs (CORs) and heliospheric imagers (HIs). For estimating the initial 3D kinematics of the CMEs, we reconstruct them in COR field of view by applying the Graduated Cylindrical Shell (GCS) forward fitting model (Thernisien et al. 2009) on the images obtained from *STEREO*/COR and *SOHO*/LASCO. We have attempted to fit the diffuse front of the CMEs which seems to envelope the loop front. The diffuse front is often formed due to local density compression at a wave front while loop front is formed due to transported piled-up plasma at the outer boundary of the flux rope. We note that the most difficult and important part in fitting halo CMEs is not fitting the two *STEREO* views but fitting the LASCO view. Further, the evolution of the CMEs is examined in the heliosphere by carefully examining the sequence of running and base difference HIs images. It is noticed that the CME signals are not sufficient to track the specific features in the sequences of images. Therefore, we constructed the time-elongation maps, conventionally called as *J*-maps, (Sheeley et al. 2008; Davies et al. 2009) using the running difference images of HI-1 and HI-2. The tracking and measuring of the time-elongation profiles of the evolving CMEs are carried out from the *J*-maps.

**Table 1.** Selected CMEs Events

| Events | *STEREO* Observations | Collision sites | collision phase | Accuracy |
|---|---|---|---|---|
| 2011 February 14-15 | Both A & B | 24 $R_\odot$ | well identified | Highest |
| 2012 June 13-14 | Both A & B | 100 $R_\odot$ | well identified | Highest |
| 2010 May 23-24 | Both A & B | 42 $R_\odot$ | end phase poorly identified | moderate |
| 2012 March 4-5 | Both A & B | 160 $R_\odot$ | well identified | moderate |
| 2012 November 9-10 | only A | 30 $R_\odot$ | well identified | moderate |
| 2013 October 25 | only B | 37 $R_\odot$ | well identified | moderate |
| 2011 August 3-4 | Both A & B | 145 $R_\odot$ | end phase not identified | Lowest |
| 2012 September 25-28 | Only A | 170 $R_\odot$ | well identified | Lowest |

NOTE—From the left: the first, second, third, fourth and fifth columns show the date of events, availability of observations from *STEREO* spacecraft, distance of collision site from the Sun, feasibility of marking the boundaries of collision phase, and accuracy assigned for the analysis, respectively. The estimation of collision site is made from the derived kinematics of the colliding CMEs as described in Section 2.1.1 to 2.1.8 .

Thereafter, an appropriate reconstruction method is applied on time-elongation profiles to estimate the 3D kinematics of the CMEs, which will be further used to identify the collision site, duration of collision, as well as pre- and post-collision dynamics. In light of the earlier studies regarding relative performance of reconstruction methods (Lugaz 2010; Liu et al. 2013; Mishra et al. 2014), we use the stereoscopic self-similar expansion (SSSE) (Davies et al. 2013) or self-similar expansion (SSE) (Davies et al. 2012) method to the *J*-maps of the CMEs. The SSSE method is used for the CMEs observed in HI field of view of both *STEREO-A* and *B* otherwise SSE method is used when the CMEs were observed either in HI-A or HI-B only.

To implement SSE or SSSE methods, an input of the appropriate value of cross-sectional angular half-width ($\lambda$) of the CME is required. For SSE method, an additional input of propagation direction of the CME is required. These inputs are obtained by applying GCS model to the CMEs in the COR field of view. It has been highlighted that use of different value of $\lambda$ with SSE or SSSE methods gives different estimate of the kinematics of the CMEs propagating away from the observer (Liu et al. 2013; Mishra & Srivastava 2015). Earlier studies have found that for CMEs propagating towards the Earth with *STEREO* behind the Sun, SSE or SSSE method should be implemented with $\lambda$ value of 90° (Liu et al. 2013, 2014; Mishra et al. 2015b; Vemareddy & Mishra 2015). The error in the kinematics from the methods applied and the difference in the estimated direction of the CMEs in COR and HI field of view will be discussed in Section 5. In the following sections, we will describe the tracking, pre- and post-collision kinematics and mass of the selected CMEs. The description of the selected cases is arranged sequentially in our study, considering the date of the events in ascending order and their assigned accuracy in descending order as per Table 1.

### 2.1.1. *2011 February 14-15 CMEs*

The CME launched on 2011 February 14 (hereinafter, CME1) and February 15 (hereinafter, CME2) have been analyzed before focusing on the kinematics, related forbush decrease (Maričić et al. 2014), their interaction corresponding to different position angles (Temmer et al. 2014), their geometrical properties and the coefficient of restitution for the head-on collision scenario (Mishra & Srivastava 2014). Our present study focuses on the nature of collision in the oblique collision scenario and the uncertainties involved therein. The parameters for CME1 and CME2 for the best visual GCS fitting (Figure 2 of Temmer et al. 2014) are listed in Table 2. The 3D speeds of CME1 and CME2 are noted as 420 km s$^{-1}$ and 580 km s$^{-1}$, respectively. The kinematics of CME1 and CME2 suggests their possible collision at some location in the heliosphere.

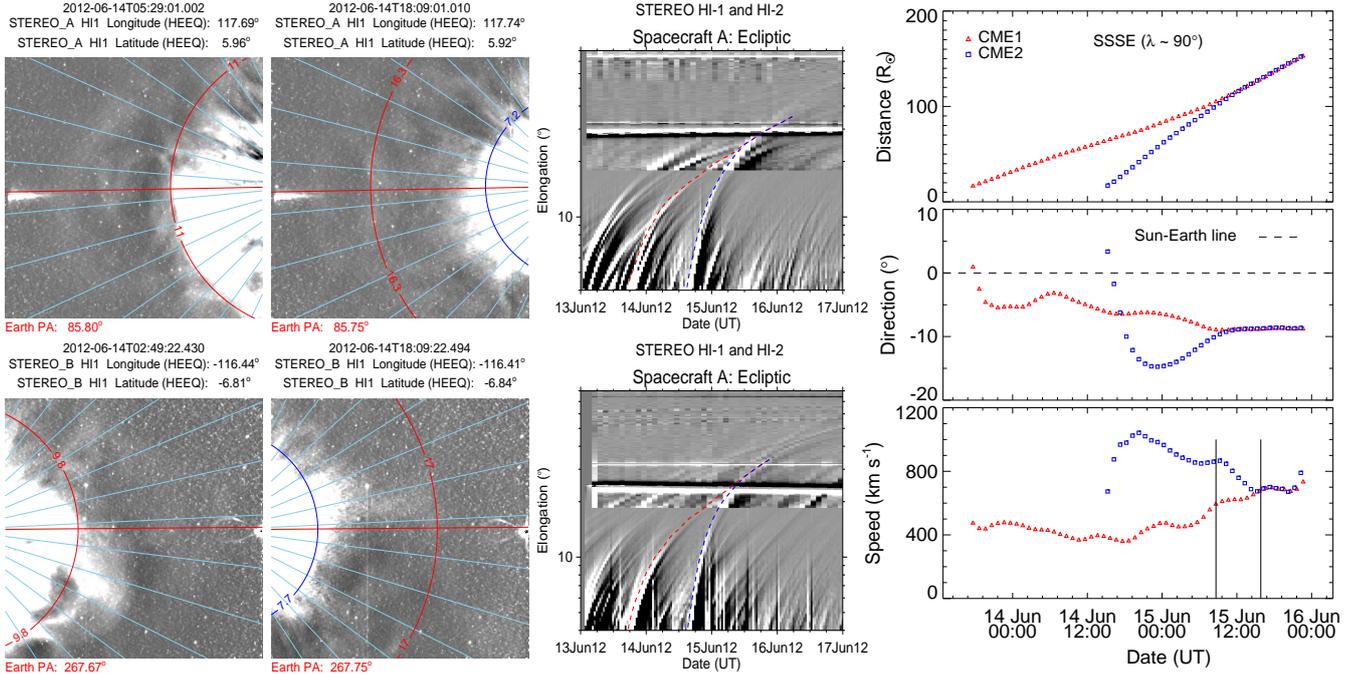

**Figure 1.** Left panel: In the top, the figures from left to right show the HI1-A base difference images at two different time and *J*-map constructed using the running difference images of HI1 and HI2. The bottom panel shows the same as the top but using HI-B images. The derived elongation of the CME1 (with red) and CME2 (with blue) are overplotted on the HI images and *J*-maps. Right panel: from the top, the first, second and third panels show the variation in distance, direction and speed of the leading edge of 2012 June CMEs. Vertical lines in the bottom panel mark the start and end of the collision phase.

The evolution of these CMEs in *J*-maps and the derived kinematics by implementing the SSSE method (Davies et al. 2013) on the time-elongation profile are respectively shown in Figures 7 and 8 of Mishra & Srivastava (2014). Based on the description of collision phase in Mishra & Srivastava (2014), we note that the collision began on 2011 February 15 at 08:25 UT and ended after 18 hr. However, the difficulty in the precise marking of start and end of momentum exchange between the CMEs is discussed in Section 5. Due to the collision, CME1 accelerated from $u_1$=300 km s$^{-1}$ to $v_1$ =600 km s$^{-1}$ and CME2 decelerated from $u_2$=525 km s$^{-1}$ to $v_2$ =400 km s$^{-1}$. We estimate the true masses of both the CMEs using COR2 images, following the method of Colaninno & Vourlidas (2009). The masses of CME1 and CME2 are estimated to be 5.4 × 10$^{12}$ kg and 4.8 × 10$^{12}$ kg, respectively. The leading edge of CME2 is around 24 $R_\odot$ and CME1 was around 26 $R_\odot$ at the beginning of the collision.

### 2.1.2. *2012 June 13-14 CMEs*

The CME of 2012 June 13 (hereinafter, CME1) and June 14 (hereinafter, CME2) appear to propagate southward in the COR2 images of *STEREO*, and CME2 appears more wider than CME1. We have applied the GCS forward fitting model to the contemporaneous images of the CMEs obtained from the SECCHI/COR2-B, *SOHO*/LASCO-C3 and SECCHI/COR2-A coronagraphs. We find the propagation direction of CME1 along E15S26 at 13.5 $R_\odot$. The propagation direction for the following CME2 was along E02S31 at 14.2 $R_\odot$. In addition to the propagation directions, GCS derived parameters for the CMEs are listed in Table 2. Around 14 $R_\odot$, the 3D speed of CME1 is noted as 560 km s$^{-1}$ and for CME2 it is 900 km s$^{-1}$. The directions and speeds of the CMEs suggest that they possibly collide during the heliospheric evolution.

These CMEs were well observed in the HI-A and HI-B field of view of *STEREO*. The base-difference HI images and the constructed *J*-maps revealing the kinematic evolution of these CMEs are shown in Figure 1. The base image used here is the minimum background created from a sequence of HI images. The tracked features come in close contact with each other and appear to merge around 25° elongation and can be tracked further up to 35°. The kinematics obtained from implementing SSSE method on the derived time-elongation profiles of these CMEs are shown in the right panels of Figure 1. The collision began on 2012 June 15 at 08:38 and continued for 7.2 hr. At the beginning of the collision, the leading edge of CME2 was at 100 $R_\odot$ and that of CME1 at 105 $R_\odot$. During the collision, they traveled a distance of around 25 $R_\odot$ before reaching to an approximately equal speed. The collision led to an acceleration of the preceding CME1 from 590 km s$^{-1}$ to 680 km s$^{-1}$ and a deceleration of the following CME2 from 865 km s$^{-1}$ to 680 km s$^{-1}$. The masses of CME1 and CME2 are estimated to be $8.4 \times 10^{12}$ kg and $9.2 \times 10^{12}$ kg, respectively.

### 2.1.3. *2010 May 23-24 CMEs*

The analysis of CMEs of 2010 May 23 (hereinafter, CME1) and May 24 (hereinafter, CME2) have been reported by Lugaz et al. (2012). The GCS derived parameters (Figure 3 of Lugaz et al. 2012) of these CMEs are listed in the Table 2. The speeds of CME1 and CME2 at the outer edge of COR field of view are estimated as 450 km s$^{-1}$ and 650 km s$^{-1}$, respectively. Figure 4 and Figure 6 of Lugaz et al. (2012) show the *J*-maps constructed from HI images and derived kinematics for these CMEs, respectively. A big data gap from *STEREO-B* just after the beginning of the collision prevents to accurately mark the collision phase, however the data from *STEREO-A* serves our purpose with limited accuracy. It is noted that the leading edge of CME2 collided with the back of magnetic ejecta of CME1 around 00:09 on May 25. Based on the analysis, we find that collision duration is as short as only 2.5 hr wherein large errors are expected (Lugaz et al. 2012). At the beginning of the collision, the leading edges of CME2 and CME1 were around 42 $R_\odot$ and 65 $R_\odot$ from the Sun. The collision led to an acceleration of CME1 from 360 km s$^{-1}$ to 420 km s$^{-1}$ and a deceleration of CME2 from 600 km s$^{-1}$ to 380 km s$^{-1}$. The masses of CME1 and CME2 in COR field of view exploiting the multi-viewpoint of *STEREO* are measured as $4.5 \times 10^{12}$ kg and $2.2 \times 10^{12}$ kg, respectively.

### 2.1.4. *2012 March 4-5 CMEs*

The GCS derived parameters for CME of 2012 March 4 (hereinafter, CME1) and March 5 (hereinafter, CME2) are put in Table 2. The 3D speeds of the CME1 and CME2 were around 1025 km s$^{-1}$ and 1300 km s$^{-1}$ at 16.5 $R_\odot$ and 10.7 $R_\odot$ from the Sun, respectively. The base-difference HI images, constructed *J*-maps and the kinematic evolution of these CMEs are shown in Figure 2. The kinematics derived from SSSE method show the signature of collision as an exchange of momentum between CME1 and CME2. The commencement of collision is marked at 07:12 UT when the leading edge of CME2 was at 160 $R_\odot$ and that of CME1 at 182 $R_\odot$ from the Sun. The duration of collision phase is around 4.8 hr. The exchange in the dynamics of the participating CMEs in the collision is revealed as an increase in the speed of CME1 from 475 km s$^{-1}$ to 600 km s$^{-1}$ and decrease in the speed of CME2 from 910 km s$^{-1}$ to 700 km s$^{-1}$. The masses

Table 2. Parameters for the CMEs derived from GCS model

| Events | $\phi$ (°) | $\theta$ (°) | $\alpha$ (°) | $\kappa$ | $\gamma$ (°) | $h_f$ ($R_\odot$) | $\omega_{EO}/2$ (°) |
|---|---|---|---|---|---|---|---|
| Feb 14 at 18:24 UT (CME1) | 6 | 4 | 32 | 0.28 | -8 | 10 | 16 |
| Feb 15 at 02:24 UT (CME2) | -3 | -11 | 18 | 0.37 | 25 | 11 | 22 |
| Jun 13 at 13:25 UT (CME1) | -15 | -26 | 20 | 0.55 | -64 | 13.5 | 33 |
| Jun 14 at 14:12 UT (CME2) | -2 | -31 | 31 | 0.6 | -45 | 14.2 | 37 |
| May 23 at 18:30 UT (CME1) | 12 | 6 | 23 | 0.26 | -55 | 16.3 | 15 |
| May 24 at 14:06 UT (CME2) | 26 | -5 | 15 | 0.37 | 6 | 14.5 | 22 |
| Mar 4 at 11:00 UT (CME1) | -55 | 23 | 20 | 0.6 | -36 | 16.5 | 37 |
| Mar 5 at 04:00 UT CME2) | -40 | 41 | 21 | 0.7 | -44 | 10.7 | 44 |
| Nov 9 at 15:12 UT (CME1) | 2 | -14 | 19 | 0.52 | 9 | 9.6 | 31 |
| Nov 10 at 05:12 UT (CME2) | 6 | -25 | 12 | 0.19 | 9 | 8.2 | 11 |
| Oct 25 at 08:15 UT (CME1) | -70 | 3 | 30 | 0.39 | 90 | 11.5 | 23 |
| Oct 25 at 15:15 UT(CME2) | -65 | 3 | 65 | 0.59 | 90 | 12.5 | 36 |
| Aug 3 at 14:00 UT (CME1) | 14 | 14 | 20 | 0.5 | -74 | 13 | 30 |
| Aug 4 at 04:12 UT (CME2) | 19 | 16 | 45.5 | 0.47 | 77 | 13 | 28 |
| Sep 25 at 11:24 UT (CME1) | 19 | -11 | 21 | 0.34 | 6 | 15 | 20 |
| Sep 28 at 00:12 UT (CME2) | 25 | 13 | 68 | 0.52 | -75 | 13 | 31 |

NOTE—The columns from the left to right show the time of first appearance of selected CMEs (CME1 and CME2) in LASCO-C2, longitude ($\phi$), latitude ($\theta$), half angle ($\alpha$) of conical leg of the CME, aspect ratio ($\kappa$), tilt angle ($\gamma$) around the axis of symmetry of the model, height ($h_f$) of the leading front and edge-on 3D angular half-width ($\omega_{EO}/2$) of the CME derived from implementing the GCS method of 3D reconstruction. The latitude and longitudes are given in the Stonyhurst coordinate system (Thompson 2006) in which Earth is always at the longitude of zero. The edge-on angular half-width is determined using the formulation given in Thernisien et al. (2006); Thernisien (2011). The uncertainty in propagation direction and half-angle is around $\pm 5°$, in tilt angle it is around $\pm 20°$, in an aspect ratio it is around $\pm 0.10$ and in distance it is almost $\pm 1.0$ $R_\odot$. The GCS fitted uncertainties lead to an error of $\pm 50$ km$^{-1}$ in speed values. The uncertainties in the GCS parameters are noted from the several independent attempts of applying GCS model to the CMEs.

of CME1 and CME2 are measured as $4.6 \times 10^{12}$ kg and $13.5 \times 10^{12}$ kg, respectively. We emphasize that a relatively larger ($\approx 28°$) difference in the latitude of CME1 and CME2 (Table 2) does not perfectly represent a scenario of collision occurring in ecliptic plane as assumed in our study. Our idealistic assumption would lead to different estimated values for propagation speeds, expansion speeds, and collision duration than the values actually responsible for the physical nature of collision. Such type of error is smaller for other selected cases of the CMEs.

### 2.1.5. *2012 November 9-10 CMEs*

The estimated kinematics of CME of 2012 November 9 (hereinafter, CME1) and November 10 (hereinafter, CME2) in COR and HI field of view, using multi-viewpoint observations of *STEREO*, have been reported by Mishra et al. (2015a). The obtained GCS modeled parameters (Figure 2 of Mishra et al. 2015a) of these CMEs are tabulated in Table 2. From the kinematics, we note that CME2 is relatively narrow and directed more southward than CME1. The features of CME2 were not observed well in the *STEREO-B* ecliptic *J*-map, SSSE reconstruction technique could not be implemented to estimate the CME kinematics. Figures 4, 5 and 6 in Mishra et al. (2015a) showed the constructed *J*-map, base-difference images of these CMEs with overplotted elongation, and the kinematics obtained using the SSE method. The collision began at 11:30 UT on 2012 November 10 and lasted for 5.8 hr. The leading edge of CME2 and CME1 was at around 30 $R_\odot$ and 55 $R_\odot$ from the Sun, respectively, at the commencement of the collision. The collision caused the speed of CME1 to increase from 365 km s$^{-1}$ to 450 km s$^{-1}$ while the speed of CME2 to decrease from 625 km s$^{-1}$ to 430 km s$^{-1}$. The masses of CME1 and CME2 were measured as 4.7 $\times$ 10$^{12}$ kg and 2.3 $\times$ 10$^{12}$ kg, respectively at the outer edge of COR field of view.

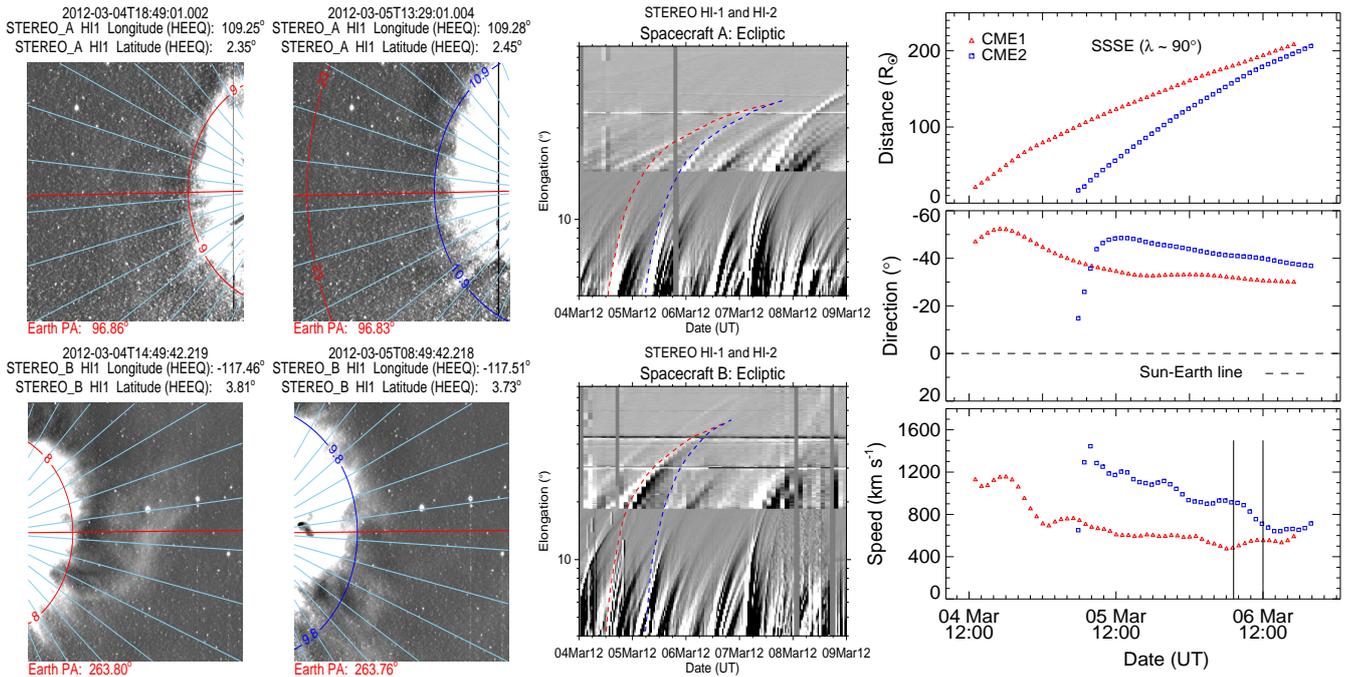

**Figure 2.** Caption is the same as for Figure 1, but this figure is for 2012 March 4-5 CMEs.

### 2.1.6. *2013 October 25 CMEs*

The subsequently launched two CMEs on 2013 October 25 are hereinafter referred as CME1 and CME2, respectively. The GCS fitted parameters for these CMEs (Figure 1 of Mishra et al. 2016) are listed in Table 2. These CMEs were propagating eastward and largely away from the Sun-earth line, therefore their leading edge could not be observed in HI-A field of view and only their flanks could be observed up to a small elongation angle. Therefore, we prefer to use only HI-B observations for our analysis. The constructed *J*-map and the overplotting of derived time-elongation from this map on the HI-B images have been shown in Figure 2 of Mishra et al. (2016), in which an obvious collision of the tracked features can be found. The SSE method is implemented to estimate the CME's kinematics (Figure 3 of Mishra et al. 2016). We note that the collision began at 23:00 UT on 2013 October 25 and lasted for 7 hr. At the beginning of the collision, the leading edge of CME2 was at around 37 $R_\odot$ and that of CME1 at around 40 $R_\odot$ from the Sun. The collision resulted in an exchange of dynamics as an acceleration of CME1 from 425 km s$^{-1}$ to 625 km s$^{-1}$ and a deceleration of CME2 from 700 km s$^{-1}$ to 500 km s$^{-1}$. The true masses of CME1 and CME2 are estimated to be 7.5 $\times$ 10$^{12}$ kg and 9.3 $\times$ 10$^{12}$ kg, respectively in the *STEREO*/COR2 field of view.

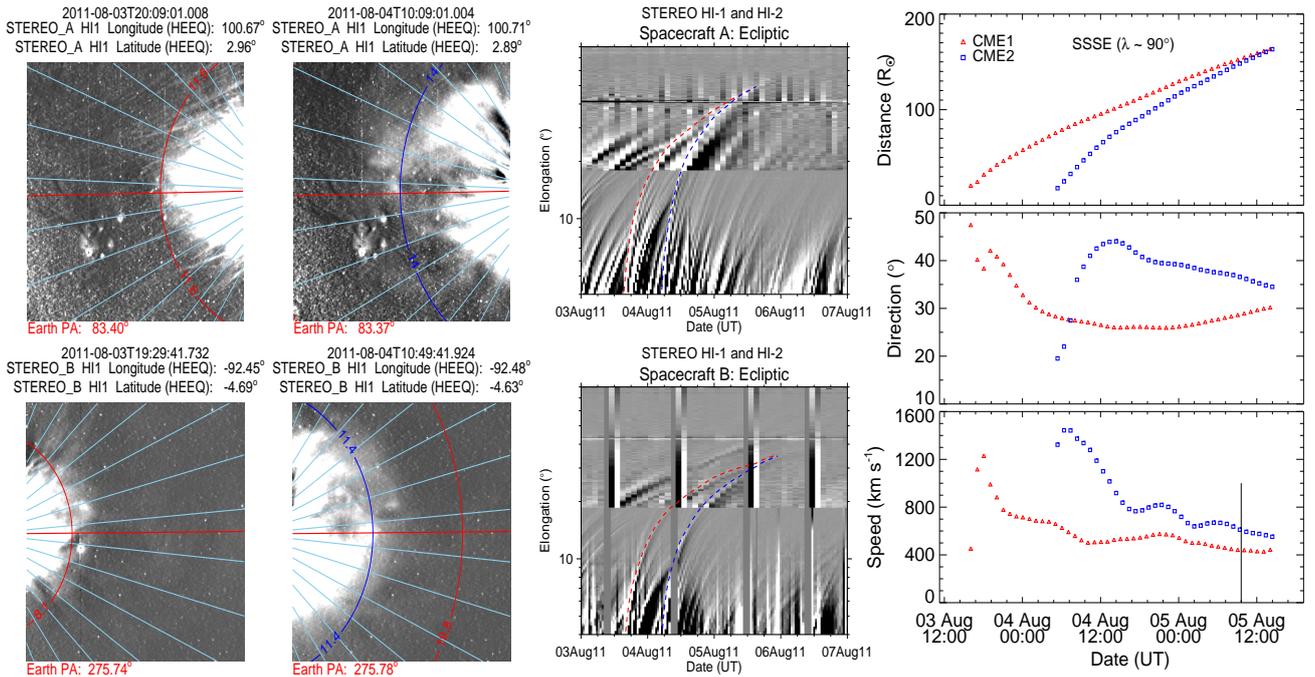

**Figure 3.** Caption is the same as for Figure 1, but this figure is for 2011 August 3-4 CMEs.

#### 2.1.7. *2011 August 3-4 CMEs*

We list the GCS fitted parameters for CMEs of 2011 August 3 (hereinafter, CME1) and August 4 in Table 2. Around 13 $R_\odot$, the speeds of CME1 and CME2 as 1100 km s$^{-1}$ and 1700 km s$^{-1}$, respectively, with their westward direction of propagation separated by only 5° to one another, ensure a collision between them. Similar to above discussed case studies, we construct the *J*-maps and then apply the SSSE method to determine the kinematics. The base-difference HI images, constructed *J*-maps and the kinematics of these CMEs are shown in Figure 3. From the kinematics, we find that the collision began at 09:35 UT on 2011 August 5 when the leading edge of CME2 was at around 145 $R_\odot$ and that of CME1 at around 150 $R_\odot$ from the Sun. Because of extremely weak signal from the CMEs even in the *J*-maps, these CMEs could not be tracked longer and therefore the end phase of the collision could not be marked. The collision occurred near the Earth and these CMEs are found to have an equal speed in in situ observations at 1 AU. Thus, we consider the post-collision speeds of the CMEs as measured in situ at 1 AU. The uncertainties arose from this will be discussed in Section 5. The collision led to an acceleration of CME1 from 420 km s$^{-1}$ to 525 km s$^{-1}$ at the cost of decelerating CME2 from 630 km s$^{-1}$ to 525 km s$^{-1}$. The true masses of CME1 and CME2 are determined as $7.4 \times 10^{12}$ kg and $10.2 \times 10^{12}$ kg, respectively.

#### 2.1.8. *2012 September 25-28 CMEs*

The CME of 2012 September 25 (hereinafter, CME1) and September 28 (hereinafter, CME2) have been analyzed in depth earlier focusing on their interaction and formation of a complex ejecta resulting in a two-step geomagnetic storm (Liu et al. 2014; Mishra et al. 2015b). Figure 1 of Mishra et al. (2015b) showed the GCS fitted wire-frame on the CMEs and the fitted parameters are listed in Table 2. The signal from these CMEs in the *J*-map constructed using HI-B images are too weak to track them without ambiguity beyond 20°. Therefore, we could not implement SSSE method, instead we used the SSE method with HI-A observations. We refer to Figure 3 and 6 of Mishra et al. (2015b) for the *J*-map and obtained kinematics for

**Table 3.** CMEs parameters under oblique collision

| Events | $e$ ($\sigma$) (km s$^{-1}$) | $\Delta$KE, $\Delta p_1$, $\Delta p_2$ (%) | $u_{1c}, u_{2c}$ | $u_{12exs}$ (km s$^{-1}$) | $u_{2ex}/u_{1ex}$ | $u_{12cjr}$ (km s$^{-1}$) | $v_{21cjr}$ (km s$^{-1}$) | $\psi$ (°) | $\Delta T$ (hr) | $m_2/m_1$ | $R$ ($R_\odot$) | $e_{1D}$ ($\sigma_{1D}$) (km s$^{-1}$) |
|---|---|---|---|---|---|---|---|---|---|---|---|---|
| Feb 14-15 | 1.65 (120) | 7.3, 68, -43 | 235, 380 | 208 | 2.2 | 130 | 230 | 3.6 | 18 | 0.8 | 24 | 0.9 (142) |
| Jun 13-14 | 0.35 (40) | -1.7, 24, -15 | 380, 540 | 533 | 1.56 | 135 | 45 | 21.9 | 7.2 | 1.1 | 100 | 0 (77) |
| May 23-24 | 1.4 (15) | 1.8, 27, -30 | 285, 435 | 237 | 2.2 | 100 | 135 | 6.6 | 2.5 | 0.5 | 45 | 0.25 (43) |
| Mar 4-5 | 0 (20) | -3.5, 53, -9 | 295, 535 | 551 | 2.0 | 210 | -10 | 12.3 | 4.8 | 2.94 | 160 | 0 (224) |
| Nov 9-10 | 0 (25) | -13.4, 38, -36 | 240, 525 | 224 | 0.8 | 280 | -60 | 0.5 | 5.8 | 0.48 | 30 | 0.1 (9) |
| Oct 25 | 1.0 (50) | 0, 48, -26 | 305, 440 | 378 | 2.2 | 130 | 140 | 7.9 | 7.0 | 1.24 | 37 | 0.45 (20) |
| Aug 3-4 | 0.1 (40) | -3.7, 31, -15 | 280, 430 | 341 | 1.4 | 145 | -5.0 | 6.6 | obscure | 1.37 | 145 | 0 (24) |
| Sep 25-28 | 2.0 (30) | 3.34, 99, -13 | 285, 405 | 305 | 2.1 | 110 | 250 | 9.7 | 16.8 | 5.53 | 170 | 0.8 (120) |

Note—From the left to the right: First column shows the selected cases of the colliding CMEs. The second and thirteenth columns list the estimated value of coefficient of restitution ($e$) and the deviation ($\sigma$) determined in oblique and head-on collision scenarios, respectively. Third column lists the total change in the kinetic energy of the CMEs, change in the momentum of CME1 and CME2. The fourth column shows the pre-collision centroid speed (i.e. propagation speed) of CME1 and CME2. The fifth, sixth and seventh columns show the sum of expansion speed of the colliding CMEs, ratio of CME2 to CME1 expansion speed and relative approaching speed of the centroids of the CMEs along the line joining their centroids, respectively, at the beginning of the collision. The eighth column shows the post-collision relative separation speed of centroids of the CMEs along the line joining their centroids. The ninth, tenth, eleventh and twelfth columns show the direction of impact, duration of collision phase, the ratio of masses of CME2 to CME1, and the distance of the collision site from the Sun, respectively. The positive and negative signs show the increase and decrease in the parameters, respectively.

these CMEs. We note that the collision lead to an acceleration of CME1 from 385 km s$^{-1}$ to 710 km s$^{-1}$ and a deceleration of CME2 from 610 km s$^{-1}$ to 430 km s$^{-1}$. The collision phase began at 22:48 UT on 2012 September 29 and lasted for 16.8 hr. At the beginning of the collision, the leading edge of CME2 was at around 170 $R_\odot$ and that of CME1 at around 190 $R_\odot$ from the Sun. The true masses of these CME1 and CME2 participating in the collision have also been estimated and found to be $1.8 \times 10^{12}$ kg and $9.7 \times 10^{12}$ kg, respectively.

## 3. COEFFICIENT OF RESTITUTION OF THE CMES: ANALYSIS AND OUTCOME

The knowledge of coefficient of restitution ($e$) for the colliding CMEs may be useful for accounting few false CME arrival alarms and help to predict their arrival at the Earth more accurately. In the present study, we treat CMEs as large expanding blobs and attempt to understand their bounciness for different cases. Using the expansion speeds of CME1 and CME2 (i.e., $u_{1ex}$ and $u_{2ex}$) and their estimated leading edge speeds (i.e. $u_1$ and $u_2$) before the collision, we determined the speeds of their centroids (i.e., $u_{1c}$ and $u_{2c}$) to be used for studying collision nature. We assume that a CME expands in such a way that its angular width remains constant. It is difficult to know the true angular width of the CME approximated as a spherical bubble. The GCS model considers a CME like a hollow croissant and enables to estimate the face-on and edge-on angular widths (Thernisien et al. 2009; Thernisien 2011). The edge-on angular width of a CME basically represents the width of conical legs with which a tubular front makes the GCS structure. The edge-on half angular width is the inverse trigonometric sine function of the fitted aspect ratio ($\kappa$) of the CME from GCS model. Since the $\kappa$ represents the rate of expansion versus height of the CME, i.e. it is the ratio of the CME size at two orthogonal directions. Hence, the edge-on angular width best suits our purpose. The angular half-width ($\omega$) of the CMEs taken as edge-on angular half-width ($\omega_{EO}/2$) are listed in Table 2. We also determine the post-collision directions and speeds of centroids of the CMEs. Further, we have assumed no change in the angular widths of the CMEs before and after the collision.

We acknowledge the errors in the observed kinematics of the CMEs and possibility of their deflection during the collision. The post-collision direction of the CMEs are kept as observationally unknown parameters due to the interrelatedness of post-collision dynamics with the nature of collision. Therefore, we determined the expected (i.e., theoretical) post-collision speeds of the centroids of the CMEs ($v_{1cth}, v_{2cth}$) for a certain value of coefficient of restitution ($e$) based on the momentum conservation law. The estimated expected post-collision speeds of the centroids are converted to their corresponding leading edge speeds ($v_{1th}, v_{2th}$), which will be compared with the observed leading edge speeds ($v_1, v_2$) of the CMEs by calculating the deviation ($\sigma$) between them. After several iterations, the best suited value of $e$ of the collision of the CMEs is found at the minimum of deviation ($\sigma$). It is noted that the value of $e$ is ranges between 0 and 5 during the iteration accounting all the possibilities of the nature of collision. We also emphasize that the $\sigma$ value of up to 150 km s$^{-1}$ is satisfactory as this implies an average difference only up to 100 km s$^{-1}$ between the observed and the expected speeds of the individual CMEs. Admitting the presence of errors in tracking of the CMEs, 3D reconstruction and those raised from idealistic geometrical assumptions; an error of around $\pm 100$ km s$^{-1}$ in the speed of the CMEs is not unexpected. The mathematical formulation applied in the present study is given in Mishra et al. (2016) with details, however the core equations from there are also mentioned here in the Appendix.

### 3.1. *2011 February 14-15 CMEs*

The edge-on angular half-widths of CME1 and CME2 (i.e. $\omega_1$ and $\omega_2$) are around 16° and 22°, respectively as noted in Table 2. Considering no obvious deflection of the CMEs before the collision in HI field of view, the estimated directions (i.e. $\phi$) from GCS model in COR2 field of view (Table 2) are used for the pre-collision directions. Under a oblique collision scenario, using the estimated kinematics and the angular widths of the CMEs in Equations (2) and (3) given in Appendix, the best suited coefficient of restitution ($e$) is found to be 1.65 with the minimum value of the deviation ($\sigma$) between the observed and the expected leading edge speeds of 120 km s$^{-1}$. This leads to an increase in the momentum of CME1 by 68% and a decrease by 43% in CME2 compared to their values before the collision, and results in an increase of 7.33% in the total kinetic energy of the CMEs. The $e$ value for this collision was estimated as close to elastic ($e$=0.9) in Mishra & Srivastava (2014) under a head-on collision scenario and thus highlights the limitation of earlier studies.

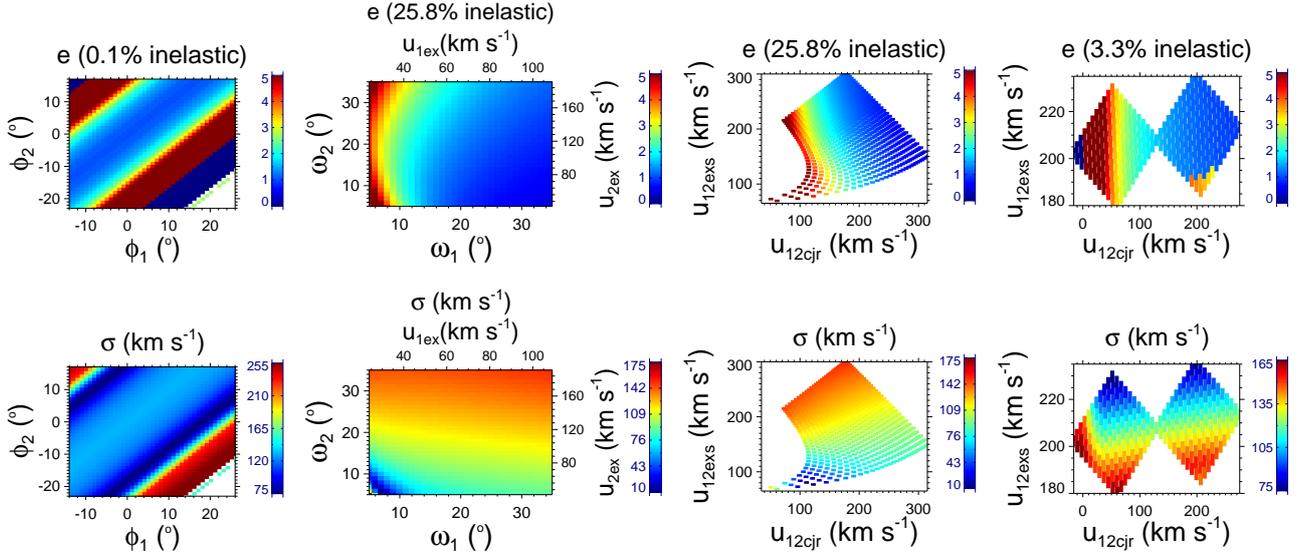

**Figure 4.** From the left: the first panel show the variation of coefficient of restitution ($e$) in the top panel and the corresponding deviation ($\sigma$) between the expected and the observed pre-collision speed in the bottom panel, for the uncertainties in the propagation direction of 2011 February CMEs. The propagation direction of CME1 and CME2 ($\phi_1$ and $\phi_1$) is shown on X and Y-axes, respectively. Second and third panel shows the variation of $e$ and $\sigma$ when the uncertainties in the angular width of the CMEs in considered. The angular half-width of CME1 and CME2 (i.e. $\omega_1$ and $\omega_2$) is shown on X and Y-axes. The expansion speed of CME1 and CME2 (i.e. $u_{1ex}$ and $u_{2ex}$) is shown at the top X-axis and right side Y-axis. Fourth panel shows the variation of $e$ and $\sigma$ when the uncertainties in initial speed of the CMEs in considered. In third and fourth panels, X and Y-axes respectively show the relative approaching speed ($u_{12cjr}$) and sum of expansion speed ($u_{12exs}$) of both the CMEs. The color bars showing the range of values corresponding to each figures are stacked.

Under the oblique collision scenario, we determined the direction of impact ($\psi$) for the collision. By direction of impact, we mean the angle between the line connecting the centroids of two colliding CMEs and the propagation velocity of CME2 relative to CME1. We also determined several parameters of the CMEs just at the beginning of the collision together with other collision parameters. Using the expansion and propagation speeds estimated from the observations of 2011 CMEs 14-15 CMEs, we have determined the ratio of CME2 to CME1 expansion speed ($u_{2ex}/u_{1ex}$), sum of their expansion speed ($u_{12exs}$), pre-collision relative approaching speed ($u_{12cjr}$) of the centroids of the CMEs, post-collision relative separation speed ($v_{21cjr}$) of the centroids of the CMEs along the line joining their centroids. These detail characteristics of the CMEs for the observed oblique collision are listed in Table 3. These parameters are also calculated for all the cases selected in our study. We attempt to compare these parameters for all the cases and examine if they show a pattern for a particular nature of collision.

In the following sections, we discuss the uncertainties in the propagation direction ($\phi$), angular size ($\omega$) and initial speed ($u$) of the CMEs and their effects on the collision nature. We keep in mind that collision condition should be satisfied while taking the uncertainties in the CME parameters. The condition requires that the speed of the leading edge of CME2 should be greater or equal to the speed of the trailing edge of CME1 along the line joining their centroids. In addition, the separation angle between the CMEs should be lesser or equal to the sum of their half-angular widths (see Appendix). Taking a range of uncertainty in observed parameters of both the CMEs, we may obtain several data points for different values of $e$, $\sigma$ and CME parameters. Based on these data points, the probability of different types of collision is determined.

### 3.1.1. *Effect of Propagation Direction*

Admitting the errors in the estimated directions of the CMEs from GCS model and the possibility of deflection of the CMEs with or without the collision, the uncertainty in the estimated value of $e$, as mentioned in Table 3, is expected. We consider an arbitrary uncertainty of $\pm 20°$ in the estimated longitude of CME1 and CME2 (i.e. $\phi_1$ and $\phi_2$). Using different pairs of longitudes, the estimated value of $e$ and the $\sigma$ is shown in top and bottom of the left panel of Figure 4. From this panel of the figure, it is clear that collision nature of these CMEs are super-elastic in nature. The values of $e$ at the top-left and bottom-right corresponds to two extreme values (i.e., 0 or 5) with large values of $\sigma$. These larger values of $\sigma$ corresponding to a larger separation angle between the CMEs suggest the lesser reliability of those $e$ values. The larger values of $\sigma$ imply that the expected dynamics of the CMEs satisfying the momentum conservation do not represent the observed collision picture. The large $\sigma$ value may be partly because of the errors in the propagation directions and the observed speeds obtained along a different direction. The probability of different nature of collision with the uncertainty in the propagation direction and the corresponding range of deviation is given in Table 4.

We also note that an increase in the error of the longitude from $\pm 1$ to $\pm 20°$, causes a decrease in the probability of a super-elastic collision from 100% to 87.7% with mean deviation of around 120 km s$^{-1}$. The increasing errors in the longitude increases the probability of perfectly inelastic (i.e. $e=0$) nature of collision from 0 to 12.2% with large value of mean deviation in speed of around 240 km s$^{-1}$. We note the exceptional 12.2% data points as unreliable causing a decrease in momentum of CME1 and a increase in momentum of CME2 (i.e., for $\Delta p_{err}$) and thus apparently violating the momentum exchange condition ($2^{nd}$ column of Table 4). All these points violating the momentum exchange condition correspond to $e=0$ ($4^{th}$ column of Table 4) and to a few of the maximum values of deviation in observed speed ($3^{rd}$ column of Table 4). We note that the uncertainty in the directions of 2011 February 14-15 CMEs causes the modification in the value of $e$. This modification would be deceptive if the larger value of deviation ($\sigma$) in the speed is overlooked.

### 3.1.2. *Effect of Angular Size*

The angular size of CME affects its expansion and centroid speed when its leading edge speed is kept constant. Using the observed kinematics as noted in Section 2.1.1, we arbitrarily consider the angular width variation between $5°$ and $35°$ and repeat the calculation of $e$. The estimated value of $e$ and $\sigma$ is shown in top and bottom of the second (from the left) panel of Figure 4. The findings of $e$, $\sigma$, range of $u_{2ex}/u_{1ex}$, and $\omega_2/\omega_1$ for a super-elastic and an inelastic collision, the percentage of data points for super-elastic collisions corresponding to the values of $u_{12exs}/u_{12cjr}$, and the percentage of data points among inelastic collisions with larger sum of expansion speed than the relative approaching speed of the CMEs are listed in Table 5.

We note that even such a large uncertainty in the angular width results in a probability of 73.2% for a super-elastic and only 25.8% for an inelastic and zero probability for a perfectly inelastic nature of collision. The bottom-right corner shows 0.47< $e$ <1 and the corresponding deviation ranges from 80 km s$^{-1}$ to 140 km s$^{-1}$. The deviation ranges between 10 km s$^{-1}$ and 175 km s$^{-1}$ for the estimated super-elastic nature of collision. The $\sigma$ is large when CME2 angular width and hence its expansion speed is larger than that of the observed value. The $e$ values for super-elastic collision correspond to the ratio of CME2 to CME1 expansion speed ranging between 0.6 and 7.9. This gives the ratio of CME2 to CME1 angular width ranging from 0.27 to 7. Among these values of $e$, around 96% have larger expansion speed of CME2 than that of CME1 expansion speed. However, $e$ values for inelastic correspond to the ratio of CME2 to CME1 expansion speed ranging between 0.35 and 1.54 and the ratio of CME2 to CME1 angular width ranging between 0.14 and 0.8. Among these values for inelastic nature, only around 47.5% have larger expansion speed for CME2 than that of CME1 expansion speed.

As suggested in earlier studies by Shen et al. (2012, 2016), we examined the characteristic of collision with relative approaching speed of the centroids of the CMEs along the line joining their centroids ($u_{12cjr}$) and the sum of their expansion speed ($u_{12exs}$) at the beginning of collision. The top and bottom of the third panel (from the left) of Figure 4 shows the variation in $e$ and $\sigma$ values. From this panel of the figure, it is clear that there is no large value of $\sigma$ for a particular nature of collision. The nature of collision tends to be super-elastic when the value of $u_{12exs}$ is getting larger than the values of $u_{12cjr}$. For instance as quoted in Table 5, among the data points with larger values of $u_{12exs}$ than their values of $u_{12cjr}$, around 84.7% points show a super-elastic ($e$>1) collision. And among the data points which have the values of $u_{12exs}$ more than 2 times the values of $u_{12cjr}$, around all the points (i.e. 100%) correspond to a super-elastic collision. But, for the data points corresponding to inelastic (0<$e$<1) nature of collision, only 39.5% have the values of $u_{12exs}$ larger than the values of $u_{12cjr}$. Thus our finding is in agreement with that previously conceptualized in Shen et al. (2012, 2016).

### 3.1.3. *Effect of Initial Speed*

We consider an uncertainty of ± 100 km s$^{-1}$ in the observed pre-collision leading edge speed of the CMEs without changing their other observed parameters. We repeat the calculation as aforementioned and the estimated value of coefficient of restitution ($e$) and deviation ($\sigma$) are shown in the top and bottom of the fourth panel (from the left) of Figure 4. The estimated values of $e$, $\sigma$, range of $u_{2ex}/u_{1ex}$ for super-elastic and inelastic nature of collisions, the percentage of data points for super-elastic collisions corresponding to the values of $u_{12exs}/u_{12cjr}$ and the percentage of data points among inelastic collisions with larger sum of expansion speed than the relative approaching speed of the CMEs, are listed in Table 6. We note the probability of 2.4% for perfectly inelastic, 3.3% for an inelastic (i.e. 0< $e$ <1), and 88.8% for a super-elastic nature of collision. The value of $\sigma$ is not specifically large for a particular type of collision and therefore the estimated vales of $e$ are reliable. Among all the data points corresponding to the values of $u_{12exs}$ larger than their values of $u_{12cjr}$, around 96.8% show super-elastic collisions. However, there are no points with 0<$e$<1 having sum of expansion speed larger than relative approaching speed.

We note the ratio of expansion speed of CME2 to CME1 ranges from 1.4 to 3.6 for $e$> 1 and 3.4 to 3.9 for 0<$e$< 1. Thus, we did not notice that a large expansion speed of CME2 give a high probability of super-elastic collisions over inelastic ones. The values of $e$ with negative approaching speed is not reliable as it is related to significantly larger value of deviation. Negative approaching speed implies that the collision of the CMEs took place because of their larger expansion speed. From the analysis of 2011 February CMEs it is obvious that a decrease in the approaching speed increases the probability of super-elastic collision.

**Table 4.** Effect of the errors of $\pm 20°$ in the observed propagation direction of the CMEs on their nature of collision

| Events | Probability for [$e=0$, $0<e<1$, $e>1$, $\Delta p_{err}$] (%) | $\sigma$ for [$e=0$, $0<e<1$, $e>1$, $\Delta p_{err}$] (km s$^{-1}$) | $e$ for $\Delta p_{err}$ |
|---|---|---|---|
| Feb 14-15 | 0, 0.1, 87.7, 12.2 | NA, NA, 75-230, 235-255 | 0 |
| Jun 13-14 | 0, 65.1, 21.7, 13.2 | NA, 35-45, 25-100, 125-230 | 0 |
| May 23-24 | 0, 40.8, 39.5, 19.7 | NA, 5-115, 10-125, 140-160 | 0 |
| Mar 4-5 | 61.8, 23.7, 10.3, 4.2 | 1-45, 0-5, 5-175, 150-215 | 0 & 5 |
| Nov 9-10 | 48.3, 34.3, 16, 1.4 | 10-175, 1-15, 20-150, 160-175 | 0 |
| Oct 25 | 0, 15.1, 67.2, 8.9 | NA, 50-52, 25-165, 180-250 | 0 |
| Aug 3-4 | 0, 76.6, 18.8, 4.6 | NA, 30-40, 20-110, 115-170 | 0 |
| Sep 25-28 | 0, 0, 89.2, 10.8 | NA, NA, 30-290, 260-310 | 0 & 5 |

NOTE—The probability of the nature of collision due to an uncertainty of $\pm 20°$ in the observed propagation direction ($\phi$) of the CMEs. The first column shows the selected cases of colliding CMEs. The second and third columns respectively show the probability of different nature of collisions (perfectly inelastic as $e=0$, inelastic as $0<e<1$, super-elastic as $e>1$ and erroneous momentum exchange as $\Delta p_{err}$) and the corresponding range of the deviation ($\sigma$) values. The $\Delta p_{err}$ stands for the scenario of an unexpected decrease in momentum of CME1 and an increase in momentum of CME2. The values of coefficient of restitution ($e$) corresponding to the points of incorrect momentum exchange ($\Delta p_{err}$) is noted in fourth column.

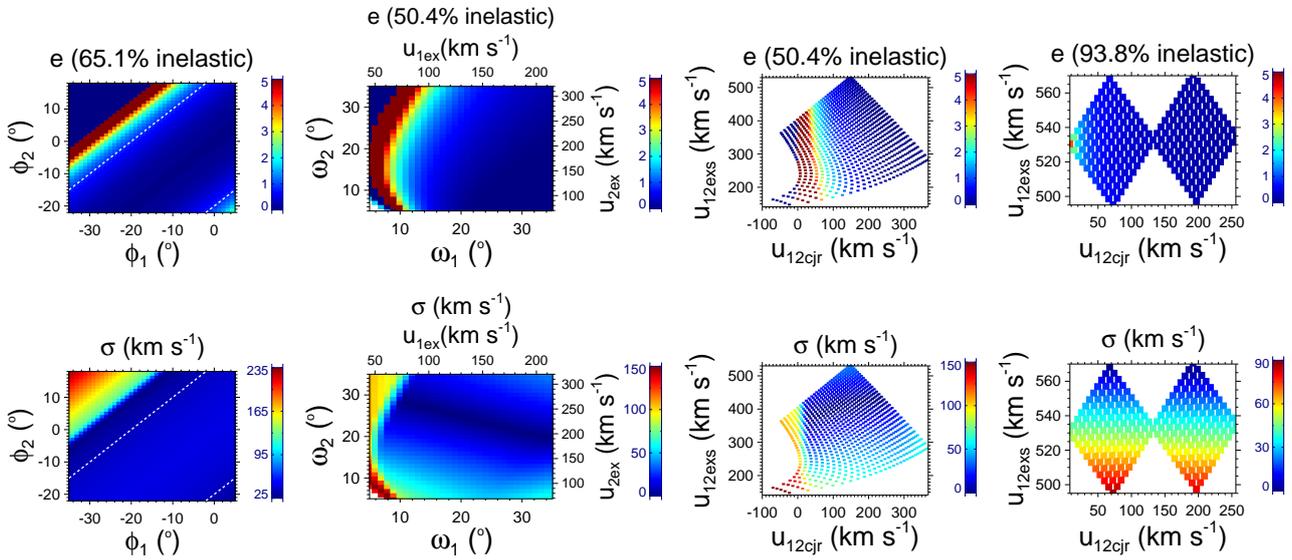

**Figure 5.** Caption is the same as for Figure 4, but this figure is for 2012 June 13-14 CMEs. The white dashed lines mark the region of $0<e<1$.

From the analysis, we found that even with the large uncertainties chosen in the directions, sizes and speeds of the CMEs, the most probable nature of collision is super-elastic for the CMEs of 2011 February.

### 3.2. *2012 June 13-14*

The estimated coefficient of restitution ($e$) and the corresponding parameters for the CMEs of 2012 June 13-14 participating in the observed oblique collision scenario is listed in Table 3. The value of $e$ is found to be 0.35 with $\sigma$ of 40 km s$^{-1}$. However, in head-on collision scenario the value of $e$ is noted as zero. The nature of collision is understood as inelastic which caused a decrease in the total kinetic energy of the CMEs by 1.7%, an increase in the momentum of CME1 by 24% and a decrease in the momentum of CME2 by 15% compared to their values before the collision. The analysis for assessing the uncertainties in $e$ is done in the similar manner as for 2011 February 14-15 CMEs described in Section 3.1. The results obtained due to the uncertainties in propagation directions, angular half-widths and speeds are given in Table 4, 5 and 6, respectively, and shown in the Figure 5. Increasing the uncertainties in the propagation directions up to $\pm 20°$ lead to an increase in the probability by around 13.2% for perfectly inelastic and around 21.7% for super-elastic. The larger probability of around 65.1% for inelastic (0<$e$<1) nature of collision with smaller value of $\sigma$ between 35 and 45 km s$^{-1}$ over other nature of collision. The data points for perfectly inelastic collision violate the momentum exchange condition and also give larger values of $\sigma$ (i.e., 125-230 km s$^{-1}$), thus becomes unreliable.

Table 5 and the second panel from the left in Figure 5 also give the preference for inelastic nature of collision. We note that data points for super-elastic collisions give the ratio of CME2 to CME1 expansion speed (angular width) ranging between 0.64 and 4.8 (0.38 and 4.2) which is larger than the values for inelastic nature of collision. Third panel of the figure shows an increase in the probability of a super-elastic collision with a decrease in the relative approaching speed ($u_{12cjr}$). There are 31.6% points with $e$>1 among the points having sum of expansion speed ($u_{12exs}$) of the CMEs larger than their relative approaching speed ($u_{12cjr}$). The probability of super-elastic collision increases with increasing the ratio of $u_{12exs}$ to $u_{12cjr}$ of the CMEs. For instance, there are around 76% points with $e$>1 among the points having the values of $u_{12exs}$ as five times larger than their values of $u_{12cjr}$. However, around 98% points among all the points for 0<$e$<1 have larger value of $u_{12exs}$ than the values of $u_{12cjr}$. There is a zero probability to have the values 0<$e$<1 with $u_{12exs}$ having six times greater than $u_{12cjr}$. The uncertainty in the speed also gives larger probability for inelastic nature of collisions with typically smaller value of $\sigma$ (fourth panel of Figure 5). However, no significant difference between the ratio of CME2 to CME1 expansion speed is noted for super-elastic and inelastic nature (Table 6). There are 6% points with $e$>1 among the points having the values of $u_{12exs}$ larger than their $u_{12cjr}$. Further, among the points having the values of $u_{12exs}$ as fifteen times larger than their values of $u_{12cjr}$, all the points have $e$>1. Carefully looking the values noted in the second row of Table 4, 5 and 6, we decide the nature of collision to be inelastic for the CMEs of 2012 June 13-14.

### 3.3. *2010 May 23-24*

Using the observed CMEs parameters in oblique collision scenario, the value of $e$ is 1.4 and the corresponding change in energy and momentum of the CMEs are listed in third row of Table 3. The value of $e$ in head-on collision scenario is 0.25 which is largely underestimated. The effect of uncertainties in propagation directions, angular half-widths and initial speeds on the collision characteristics is shown in Figure 6 and listed in third row of Table 4, 5 and 6, respectively. The uncertainties in propagation directions up to $\pm 20°$ lead to a decrease in the probability of super-elastic collision from 100% to 39.5%, and an increase of inelastic collision from 0 to 40.8% and of perfectly inelastic collision from 0% to 19.7%. Perfectly inelastic collision is not reliable as it violates the momentum condition and give large value of $\sigma$.

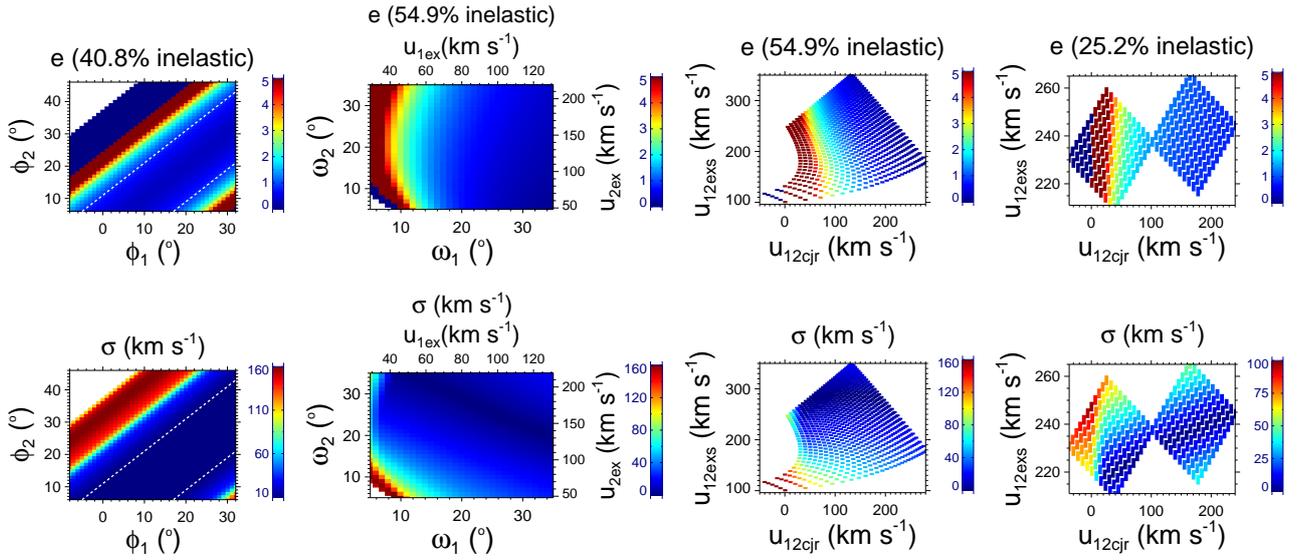

**Figure 6.** Caption is the same as for Figure 4, but this figure is for 2010 May 23-24 CMEs.

Table 5 shows a larger probability of an inelastic collision for these CMEs. From the table, we note that the ratio of CME2 to CME1 expansion speed ($u_{2ex}/u_{1ex}$) and angular width ($\omega_2/(\omega_1)$) is significantly larger for $e>1$ than for $0<e<1$. Based on the number of data points, the probability of $e>1$ increases from 48.9% to 92.1% as the ratio of $u_{12exs}$ to $u_{12cjr}$ increases from greater than 1 to 4. From the third and fourth panel of Figure 6, it is clear that low approaching speed gives more data points for $e>1$. However, the values of $e$ with negative approaching speed give relatively larger values of $\sigma$ and thus are not reliable. The errors in the initial speeds of the CMEs give a probability of around 60.7% for super-elastic collision, however this corresponds to a maximum value of $\sigma$ as around two times (i.e., 90 km s$^{-1}$) than that obtained (i.e., 40 km s$^{-1}$) for inelastic collisions. Thus, for these CMEs the nature of collision is very sensitive to the speeds measurements. As the values of $u_{12exs}/u_{12cjr}$ increased from 1 to 2, the probability of $e>1$ also increased from 60.7% to 83.5%. On taking a more dedicated study than Lugaz et al. (2012) for estimation of $e$ value, we decide that the nature of collision of the CMEs of 2010 May 23-24 may vary from inelastic to super-elastic under the involved uncertainties in their kinematic parameters.

### 3.4. *2012 March 4-5*

The value of $e$ for these CMEs of 2012 March 4-5 is estimated as zero under the oblique collision scenario, and implies 100% probability for perfectly inelastic collision. The observed characteristics of the CMEs at the beginning of the collision and the details of collision are given in Table 3. The uncertainties of $\pm 20°$ in the observed propagation directions still give a larger probability of around 65.8% for perfectly inelastic nature of collision over any other nature (first panel of Figure 7). However, there are 4.2% data points which violate the momentum exchange condition (i.e., for $\Delta p_{err}$) and give extreme values of $e$ (i.e., either 0 or 5) with higher value of $\sigma$, and therefore they are unreliable (Table 4).

The errors in the angular half-width of the CMEs also suggest that probable nature of collision is perfectly inelastic. Similar to the other cases described above, the ratio of CME2 to CME1 expansion speed is significantly higher for the values of $e>1$ than that for $e<1$ (second panel of Figure 7). In the different samples of data points having the values of $u_{12exs}/u_{12cjr}$ as greater than from 1 to 4, the percentage of data

Table 5. Effect of the Errors in the Angular Size (5° - 35°) of the CMEs on their Nature of Collision

| Events | Probability for [e=0, 0<e<1, e>1] (%) | σ for [e=0, 0<e<1, e>1] (km s$^{-1}$) | $u_{2ex}/u_{1ex}$ ($\omega_2/\omega_1$) for e>1 | $u_{2ex}/u_{1ex}$ ($\omega_2/\omega_1$) for 0<e<1 | e>1 among $\frac{u_{12exs}}{u_{12cjr}}>1$ (%) | e>1 among $\frac{u_{12exs}}{u_{12cjr}}>2$ (%) | 0<e<1 with $\frac{u_{12exs}}{u_{12cjr}}>1$ (%) |
|---|---|---|---|---|---|---|---|
| Feb 14-15 | NA, 25.8, 73.2 | NA, 80-140, 10-175 | 0.60-7.9 (0.27-7) | 0.38-1.54 (0.14-0.8) | 84.7 | 100 | 39.5 |
| Jun 13-14 | 19.7, 50.4, 29.7 | 0-150, 0-70, 0-145 | 0.64-4.8 (0.38-4.2) | 0.44-2.0 (0.23-1.6) | 31.6 | 42.8 | 97.9 |
| May 23-24 | 1.0, 54.9, 43.6 | 155-160, 0-150, 0-160 | 0.56-7.6 (0.28-7) | 0.36-2.8 (0.14-1.8) | 48.9 | 64 | 78.7 |
| Mar 4-5 | 65, 26, 8.8 | 25-205, 25-125, 30-160 | 0.89-8.7 (0.4-7) | 0.74-5.6 (0.33-4.1) | 11.1 | 24.7 | 97.5 |
| Nov 9-10 | 40.7, 58, 1.3 | 0-50, 0-45, 25-30 | 5.6-7.8 (4.9-7.0) | 0.75-6.8 (0.38-5.6) | 2 | 5.8 | 78.4 |
| Oct 25 | 0.8, 74.9, 23.8 | 15-20, 0-55, 0-50 | 1.6-7.5 (1-7) | 0.39-2.7 (0.16-1.9) | 32.1 | 57.9 | 63.8 |
| Aug 3-4 | 37.3, 53.1, 9.4 | 0-65, 0-55, 10-55 | 2.8-6.8 (2.5-7) | 0.89-4.4 (0.56-3.6) | 11.8 | 21.2 | 92.1 |
| Sep 25-28 | 0, 39.1, 60.2 | NA, 40-85, 20-145 | 0.94-7.2 (0.56-7) | 0.35-1.44 (0.14-0.89) | 74.1 | 99.7 | 48.1 |

NOTE—The probability of the nature of collision due to a varying 3D edge-on angular half-width of the CMEs (i.e., $\omega_1$ and $\omega_2$) between 5° and 35°. From the left, the first, second and third columns show the selected cases of colliding CMEs, probability of different nature of collisions and the corresponding range of the deviation ($\sigma$) values, respectively. Fourth and fifth columns show the ratio of expansion speed (angular width) of CME2 to CME1 for the points having e>1 and for the points with 0<e<1, respectively. Sixth [Seventh] columns show the percentage of points having e> 1 among the points which correspond to sum of expansion speed ($u_{12exs}$) of the CME1 and CME2 greater [two times greater] than the relative approaching speed of the centroids ($u_{12cjr}$) of the CMEs. The eighth column shows the percentage of data points among 0<e< 1 with the values of $u_{12exs}$ greater than the values of $u_{12cjr}$ of the CMEs.

**Table 6.** Effect of the Errors of $\pm 100$ km s$^{-1}$ in the Observed Speed of the CMEs on their Nature of Collision

| Events | Probability for [$e=0$, $0<e<1$, $e>1$] (%) | $\sigma$ for [$e=0$, $0<e<1$, $e>1$] (km s$^{-1}$) | $u_{2ex}/u_{1ex}$ for $e>1$ | $u_{2ex}/u_{1ex}$ for $0<e<1$ | $e>1$ among $\frac{u_{12exs}}{u_{12cjr}}>1$ (%) | $0<e<1$ with $\frac{u_{12exs}}{u_{12cjr}}>1$ (%) |
|---|---|---|---|---|---|---|
| Feb 14-15 | 2.4, 3.3, 88.8 | 155-170, 110-130, 75-165 | 1.4-3.6 | 3.4-3.9 | 96.8 | 0 |
| Jun 13-14 | 0, 93.8, 6.2 | NA, 1-90, 30-65 | 1.2-1.3 | 1.2-2.0 | 6.2 | 100 |
| May 23-24 | 8.6, 25.2, 60.7 | 65-100, 0-40, 0-90 | 1.6-2.7 | 2.7-3.5 | 60.7 | 74.4 |
| Mar 4-5 | 97.9, 2.1, 0 | 15-65, 55-60, NA | NA | 1.8-1.9 | NA | 100 |
| Nov 9-10 | 100, 0, 0 | 15-75, NA, NA | NA | NA | NA | NA |
| Oct 25 | 0.4, 49.6, 49.2 | 102, 5-95, 10-95 | 1.5-2.1 | 2.2-3.2 | 49.2 | 100 |
| Aug 3-4 | 15.3, 84.3, 0.4 | 0-20, 10-90, 55 | 0.97 | 0.99-2.18 | 0.4 | 100 |
| Sep 25-28 | 0, 50.4, 48.8 | NA, 1-85, 2-95 | 1.5-2.1 | 2.2-3.3 | 48.7 | 100 |

NOTE—The probability of the nature of collision due to an uncertainty of $\pm 100$ km s$^{-1}$ in the observed pre-collision speeds of the CMEs. The first to sixth columns are the same as in Table 5. The seventh column shows the same as the eighth column in Table 5. The entry of "NA" at some places in the table refers to "Not Applicable" and the value for there has no meaning.

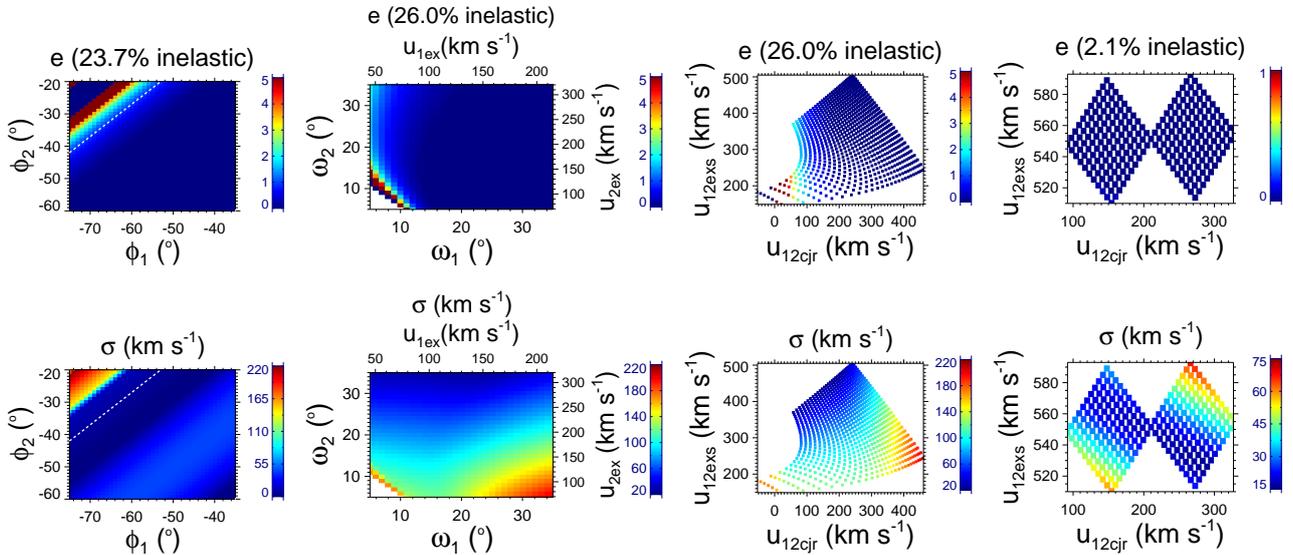

**Figure 7.** Caption is the same as for Figure 4 but this figure is for 2012 March 4-5 CMEs..

points with $e>1$ among those samples increases from 11.1% to 80.9%., i.e. the probability of $e>1$ increases with increasing the ratio of $u_{12exs}$ to $u_{12cjr}$. The errors in the initial speeds also give a significantly larger probability for perfectly inelastic collision with no chance for super-elastic collisions. In Table 6 for this case of 2012 March 4-5 CMEs, the entry corresponding to $e>1$ is made as "NA" which refers to "Not Applicable", i.e. no data points or no probability is found for $e>1$. From the speed uncertainties, there are no points having the values of 0<$e$<1 with $u_{12exs}$ greater than three times of $u_{12cjr}$. This means that as the ratio $u_{12exs}$ to $u_{12cjr}$ becomes greater than from 1 to 3, the probability of inelastic nature of collisions decreases from 100% to 0%. Despite taking the uncertainties as listed in fourth row of Table 4, 5, and 6, we always note a smaller probability for inelastic (0% to 25.9%) or super-elastic (0 to 10.5%) collisions, and therefore the collision nature of 2012 March 4-5 CMEs is decided as perfectly inelastic.

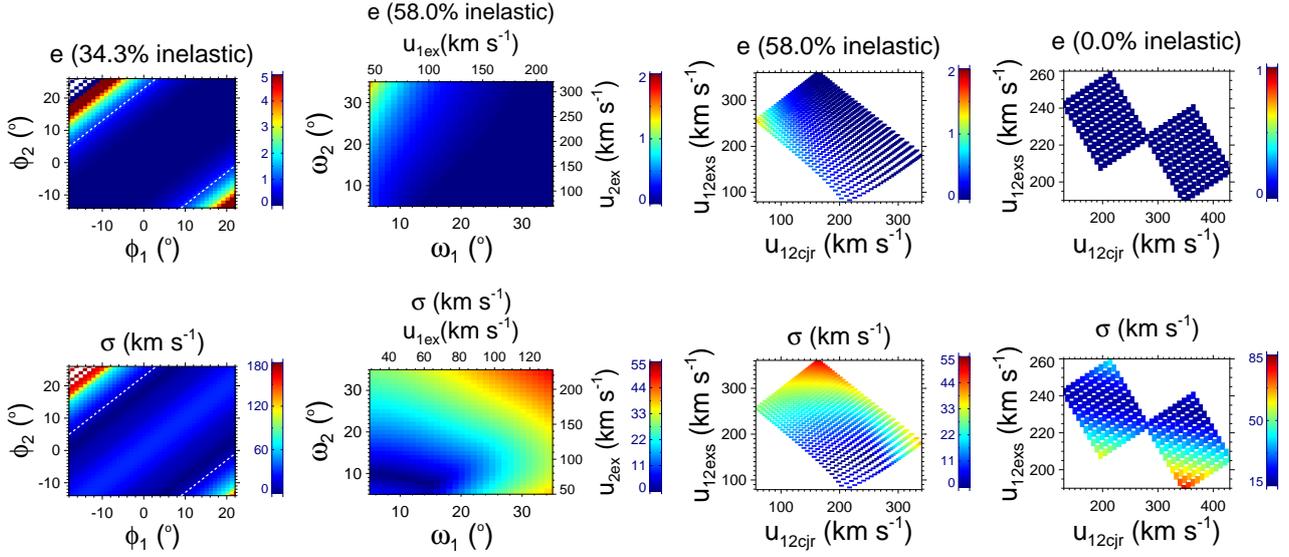

**Figure 8.** Caption is the same as for Figure 4 but this figure is for 2012 November 9-10 CMEs.

### 3.5. *2012 November 9-10*

The collision characteristics of 2012 November CMEs are mentioned in Table 3 which suggest for perfectly inelastic collision. The values of $e$ and $\sigma$ estimated taking the uncertainties in the CMEs parameters are shown in Figure 8. This is consistent with the study by Mishra et al. (2015a) where they considered head-on collision scenario. The probability of different nature of collision and several characteristics of the CMEs due to the uncertainties in their propagation directions, angular sizes, and initial speeds are given in Table 4, 5 and 6. From the tables, we note that the uncertainties of ±20° in propagation directions causes a decrease in the probability of perfectly inelastic collision up to 48.3%, however this remains higher over any other nature.

Under the uncertainties in the angular widths, we have several data points for $e$ corresponding to different pairs of the widths. We note that in the different samples of data points having the values of $u_{12exs}/u_{12cjr}$ as greater than from 1 to 4, the probability of having $e$>1 (i.e., super-elastic collisions) among those samples increases from 2% to 100%. From Table 5, the probability of inelastic collision is 58% and among this around one third correspond to $e$ values less than 0.1 and the probability of perfectly inelastic collision is 40.7%. The inelastic and super-elastic values of $e$ comes for very smaller value of angular width of CME1 than that of CME2. This is in contrast to the observations where CME1 angular half-width is around three times of CME2 angular half-width. Taking the errors in the speeds also give the probability of 100% for perfectly inelastic collision. It is also clear from the third panel of Figure 8 that low approaching speed leads to a larger probability for super-elastic collision. Carefully looking the variations of the $e$ and $\sigma$ values together with keeping in mind the observed CMEs parameters, we attribute the nature of collision for the CMEs of 2012 November 9-10 to be perfectly inelastic.

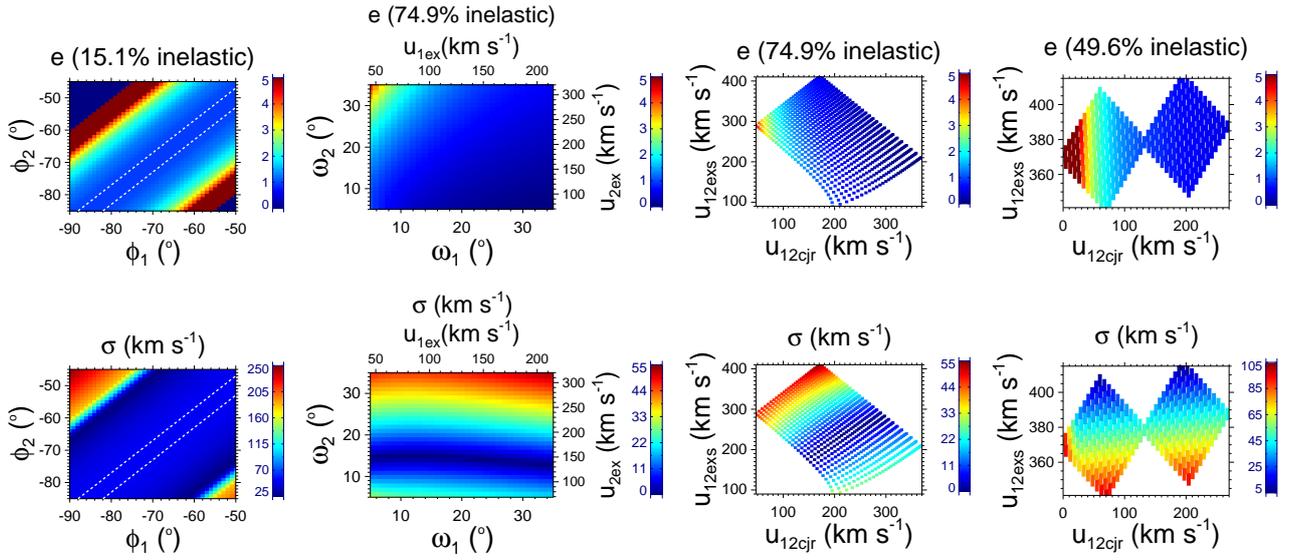

**Figure 9.** Caption is the same as for Figure 4 but this figure is for 2013 October 25 CMEs.

### 3.6. *2013 October 25*

Under the oblique collision scenario as observed, the characteristics of 2013 October 25 CMEs at the beginning of the collision and the collision parameters are listed in Table 3. The nature of collision is found to be at the boundary of inelastic and super-elastic, i.e. perfectly elastic. This value of $e$ is slightly different with the study in Mishra et al. (2016) where a different angular width (i.e., expansion speed) of the CMEs was considered. However, similar to other analyzed cases as described above, before decisively attributing a particular nature of collision for any CMEs, we assess the uncertainty in the estimated $e$ value. Figure 9 shows the variation in $e$ and corresponding $\sigma$ value against the uncertainties in the observed CMEs directions, angular half-widths and speeds. The effect of these uncertainties on the CMEs characteristics and the collision parameters are listed in Table 4, 5 and 6. From the first panel (from the left) of Figure 9, it is clear that $e$=0 gives exceptionally larger (i.e., 180 to 250 km s$^{-1}$) value of $\sigma$ and are not reliable as they also violate the momentum exchange condition. The increase in the separation angle between the CMEs than that determined from GCS model, i.e., an increase in the longitude of one CME and a decrease in another, shifts the collision nature into the super-elastic regime with only a small probability of inelastic nature. However, the maximum value of $\sigma$ (i.e., 163) for super-elastic collisions is almost three times larger than that obtained corresponding to inelastic regime.

From Table 5, the uncertainties in the angular width of the CMEs shifts the nature of the collision into inelastic regime (with probability of 75%) to a great extent. Second panel of the figure shows (similar to other cases) that the collision tends to be a super-elastic in nature when the expansion speed of the following CME gets larger while the expansion speed of the preceding CME gets smaller. We note that the values of $u_{2ex}/u_{1ex}$ reach the maximum up to 7.5 for $e>1$ while it reaches at the maximum of 2.7 for the $0<e<1$. The probability of $e>1$ increases from 32.1% to 100% as the values of $u_{12exs}/u_{12cjr}$ becomes greater than from 1 to 3. The fourth panel of the figure and sixth row the Table 6 show that the uncertainties in the initial speeds of the CMEs give around an equal probability for an inelastic and a super-elastic collision with no obvious difference in the $\sigma$ ranging from 5 to 100 km s$^{-1}$. The values of $e$ for inelastic regime also range from 0.5 to 0.9 and thus never close to $e$=0. Further, we note that among the points where the values of $u_{12exs}$ is larger than 3 times of the values of $u_{12cjr}$, around 99.1% points show $e$>1. We decide that the probable nature of collision of 2013 October 25 CMEs may vary from inelastic to super-elastic under the reasonable uncertainties in the observed directions, widths and speeds of the CMEs.

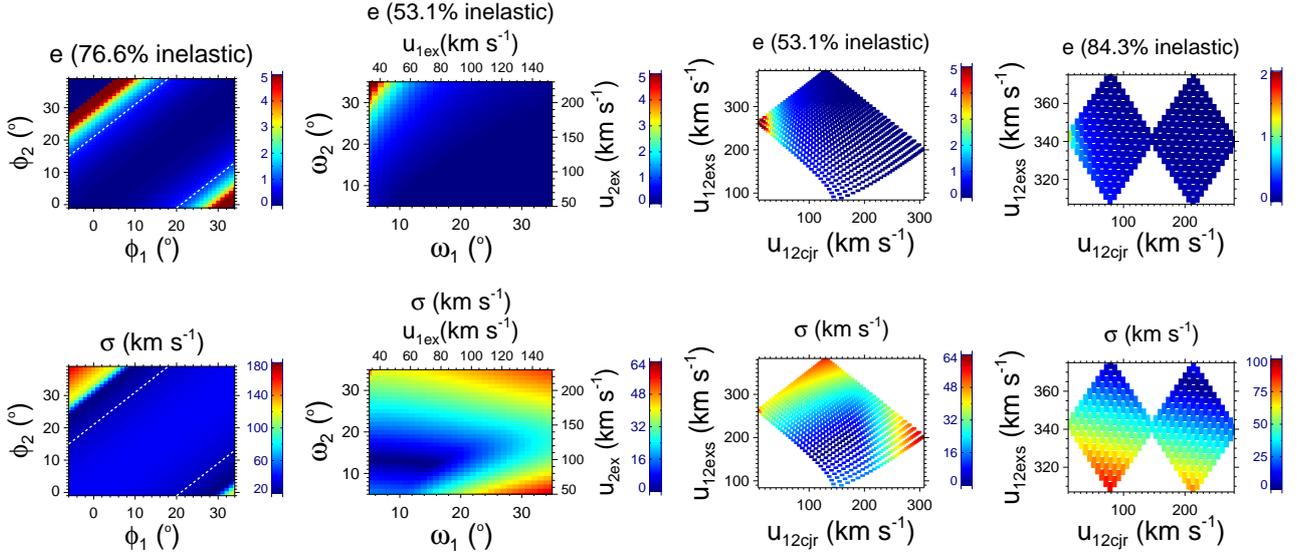

**Figure 10.** Caption is the same as for Figure 4 but this figure is for 2011 August 3-4 CMEs.

### 3.7. *2011 August 3-4*

Using the observed CMEs parameters, the value of $e$ is estimated as 0.1. The parameters derived from the propagation and expansion speeds of the CMEs and the collision parameters are noted in Table 3. The effect of uncertainties of $\pm 20°$ in the directions of the CMEs on the value of $e$ and $\sigma$ is shown in the first panel of the Figure 10 and in the seventh row of Table 4. We note that around 4.6% of data points violating the momentum exchange condition correspond to $e$=0, and therefore they are unreliable. There is a larger probability of 76.6% that collision would be inelastic with a lesser value of $\sigma$ ranging between 30 km s$^{-1}$ and 40 km s$^{-1}$.

The effect of the uncertainties in angular half-width of the CMEs is listed in Table 5 and shown in the second and third panels of Figure 10. We note that it is more likely that the values of $e$>1 correspond to a larger value of the ratio of CME2 to CME1 expansion speed before the collision than that for the values of $e$<1. We also found that among the points where the values of $u_{12exs}$ are greater than the values of $u_{12cjr}$, around 11.8% have $e$>1. And among the data points where $u_{12exs}$ is greater than six times of $u_{12cjr}$, around 100% of the points have $e$>1. This shows that a super-elastic collision is probable with a low relative approaching speed of the CMEs. Among all the data points having 0<$e$<1, around 92.1% of them have larger value of $u_{12exs}$ than $u_{12cjr}$ value. The effect of the uncertainties in speed is shown in the seventh row of Table 6 and fourth panel of Figure 10. From here we find that the probability of super-elastic collisions increases as the relative approaching speed of the CMEs decreases. There is a probability of around 84.3% for an inelastic collision with $\sigma$ value always less than 100 km s$^{-1}$, and thus it is reliable. From the results shown in the figures and the tables, we decide the nature of collision of the CMEs of 2011 August 3-4 as inelastic.

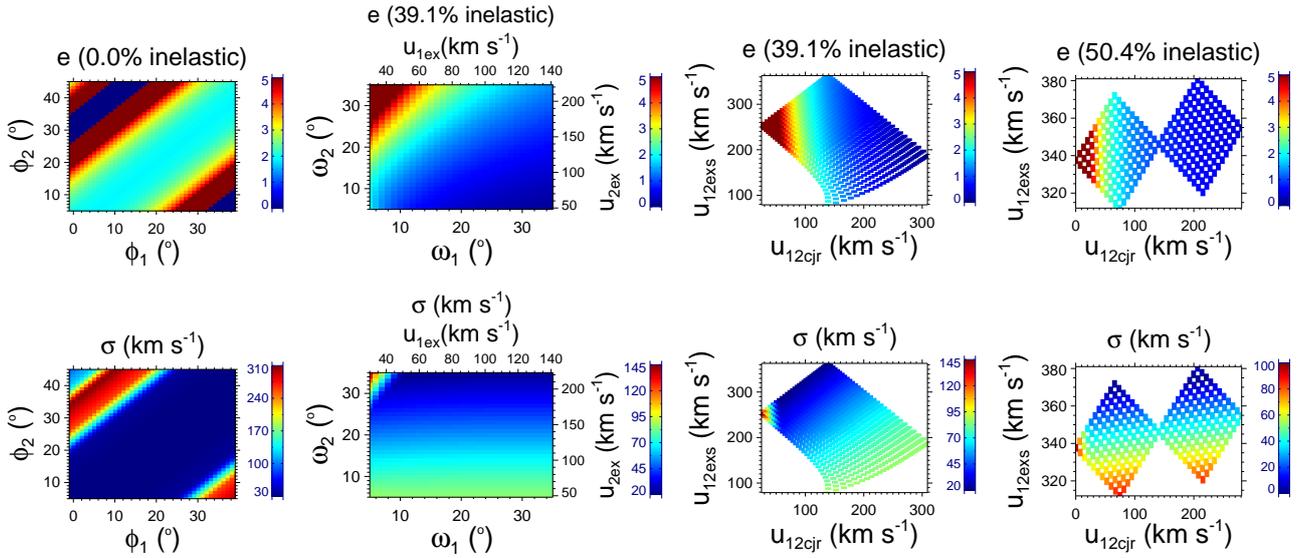

**Figure 11.** Caption is the same as for Figure 4 but this figure is for 2012 September 25-28 CMEs.

### 3.8. *2012 September 25-28*

For the CMEs of 2012 September 25 and 28, the value of $e$=2.0 is found for the oblique collision and the corresponding CMEs characteristics are listed in Table 3. Assuming the head-on collision scenario in Mishra et al. (2015b), the value of $e$ is found to be 0.8 which is obviously underestimated than that from oblique collision scenario. The effect of the errors in propagation directions, angular widths and speeds on the values of $e$ and $\sigma$ is listed in Table 4, 5 and 6, respectively. The characteristics of the CMEs derived from their expansion and propagation speeds for different nature of collision are listed in these three tables. By ignoring the points violating the momentum exchange condition and larger values of $\sigma$ while taking the uncertainties in the directions of the CMEs, we note a dominant probability of around 89.8% for a super-elastic collision (first panel of Figure 11).

Second panel of Figure 11 shows that $e>1$ is more probable with a larger expansion speed of the following CME than that of the preceding CME. For instance, the values of $u_{2ex}/u_{1ex}$ reach maximum up to 7.2 for $e>1$ while its maximum is 1.44 for $0<e<1$. Third panel of the figure shows that a decrease in the relative approaching speed ($u_{12cjr}$) of the CMEs increases the probability of a super-elastic collision. In three different samples of data points each having the ratio of $u_{12exs}$ to $u_{12cjr}$ as greater than 1, 2 and 3, the probability of $e>1$ among those samples is 74.1%, 99.7% and 100%, respectively. However, around 48.1% of the sample among the points having $0<e<1$ values are with larger values of $u_{12exs}$ than $u_{12cjr}$. The eighth row in Table 6 and the fourth panel of Figure 11 suggest for an almost equal probability for a super-elastic and an inelastic collision. However, the values of $e$ for inelastic nature of collision is always greater than 0.5. Under the reasonable errors in the directions, sizes and speeds of the CMEs of 2012 September 25-28, it is difficult to ascertain a particular collision nature as the most probable nature of collision is varying from an inelastic to a super-elastic regime.

### 4. RESULTS

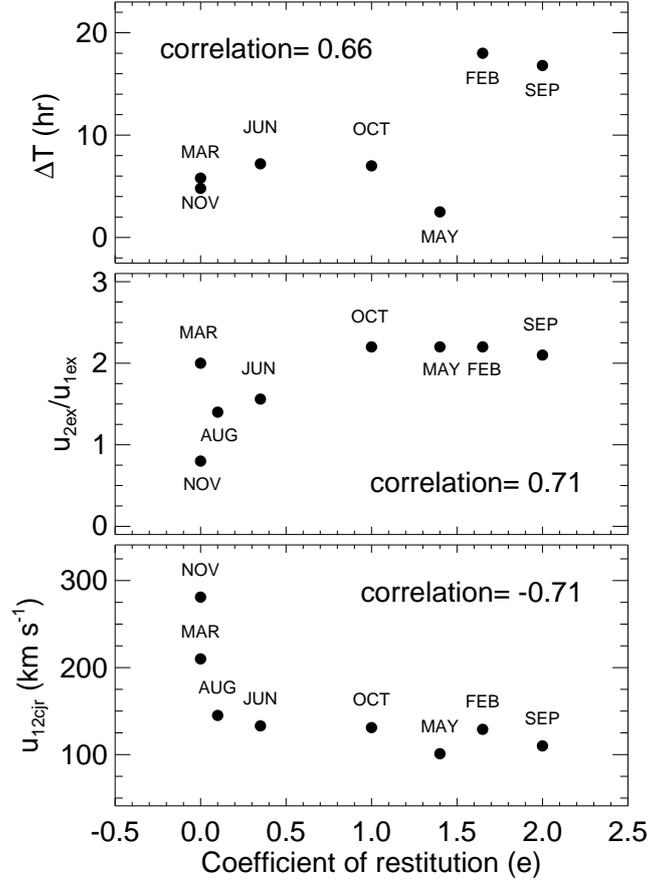

**Figure 12.** The values of coefficient of restitution is shown on the X-axis in all the panels. The collision duration, ratio of pre-collision expansion speed of CME2 to CME1 and relative approaching speed of the CMEs along the line joining their centroids are shown on the Y-axis of top, middle and bottom panels, respectively.

The results of our analysis of a total of 8 cases of the interacting CMEs are organized into two parts. The obtained physical nature of the collision using the observed parameters of the CMEs is summarized in Section 4.1. We pay attention to the relative values of CMEs parameters before and after the collision, for all the cases. In this section, we also compare different cases of the observed CMEs to find a dependence of CMEs parameters for a particular physical nature of collision (Table 3). In Section 4.2, the results from taking the uncertainties in each individual cases of the CMEs are summarized. Here we emphasize that the uncertainties in kinematic parameters of the CMEs influence the calculated value of coefficient of restitution ($e$) and thus it is a mathematical outcome (Table 4, 5 and 6). Such an analysis of uncertainties suggests that the errors in some of the parameters of the CMEs would affect the calculated value of $e$, i.e., nature of collision. Such mathematical influences on the calculated nature of collision for a selected colliding CMEs do not imply a real change in the physical nature of collision of that individual selected CMEs. Hence, the mathematical uncertainties in an individual case of the CMEs no longer represent the same physical CMEs as observed. Therefore, a mathematical treatment as done in our study do not indicate for any real change in CMEs parameters which affect the physical nature of the collision.

### 4.1. *Nature of collision: Results from the observed parameters of the CMEs*

In oblique collision scenario, we have studied several cases of the colliding CMEs using their geometrical, kinematic and mass estimates from multiple viewpoints remote observations. The important results without considering the uncertainties in measured CMEs parameters are noted in Table 3 and shown in Figure 12. On careful inspection of the figure, we notice that 4 data points of inelastic regime (e.g., Nov, Mar, Aug, and Jun) in the bottom panel of Figure 12 show a pattern of decreasing approaching speed with increasing value of $e$. Another 4 data points of elastic to super-elastic regime (e.g., 2013 October, 2010 May, 2011 February, and 2012 September) may be considered with almost a constant approaching speed with increasing value of $e$. A similar trend for data points divided in two populations can be noted for middle panel of the figure. However, unlikely to middle and bottom panels, the data points do not represent two separate populations in the top panel of the figure. All the data points of any one population could not be considered as obvious outliers as they are not unusually far from other data points. Further, all 8 data points representing 8 selected cases of interacting CMEs are not with the same precision. This is because our analysis using available single or multiple viewpoints observations have determined the CMEs parameters for different cases with different accuracy (Table 1). Also, different estimated parameters may have different precision even for a particular case of interacting CMEs. For example, as described in Section 2.1.3, the collision duration for 2010 May CMEs have a large uncertainty while we did not expect a large uncertainty for its pre-collision speed measurements. Similarly, the CMEs of 2012 March 4-5 propagating largely at different latitudes can hardly satisfy our assumptions of collision scenario in the ecliptic plane, as described in Section 2.1.4, are expected to have large errors in speed measurements while not a large errors for the collision duration. In light of aforementioned notes, we avoid using any outlier detection schemes for our extremely limited number of data points. We opine that a significant portion of the data (i.e., around 50%) cannot be considered and excluded as collective outliers, and thus we prefer to investigate each data points individually.

Among the cases studied using their observed parameters, two cases (CMEs of 2012 June 13-14 and 2011 August 3-4) show inelastic nature of collisions and two cases (CMEs of 2012 March 4-5 and 2012 November 9-10) show perfectly inelastic collision nature. One case (CMEs of 2013 October 25) shows elastic nature of collision and other three cases (CMEs of 2011 February 14-15, 2010 May 23-24 and 2012 September 25-28) show super-elastic nature of collisions. The super-elastic collision shows the collision duration as large as 18 hr for 2011 February 14-15 CMEs to as small as 2.5 hr for 2010 May 23-24 CMEs. However, the collision duration of 2010 May CMEs have large errors as it is derived using only *STEREO-A* observations due to a large data gap in *STEREO-B* just after the beginning of the collision (Lugaz et al. 2012). It is also noted that the super-elastic collisions occurs as close as 24 $R_\odot$ up to as far as 170 $R_\odot$ from the Sun. The bottom panel of Figure 12 indicates that coefficient of restitution ($e$) is negatively correlated (correlation coefficient=-0.71) with relative approaching speed. From middle panel, it is found that $e$ is positively correlated (correlation coefficient=0.71) with the ratio of CME2 to CME1 expansion speed. The top panel shows that $e$ value is positively correlated (correlation coefficient=0.66) with collision duration of interacting CMEs. We understand that a relatively better correlation could have been found if the errors in measured speeds and collision duration were smaller for 2012 March and 2010 May CMEs, respectively. Also, the data points for 2013 October, 2010 May, 2011 February and 2012 September cases indicate that the long interval of collision duration favor to some extent in determining a larger value of $e$. However, we emphasize that Figure 12 is not to show a one to one correlation between measured parameters and $e$ values. Thus, the $e$ value probably depends on several parameters and their relative contribution could not be assessed in the present study.

To understand the interrelatedness of several parameters as listed in the columns of Table 3 for all the events, we carried out principal component analysis (PCA) (Hotelling 1933; Jolliffe 2002) and the findings are put in Appendix-B. The analysis gives two significant variables as $PC_1$ and $PC_2$. A higher value of $PC_1$ comes from a combination of an increase in coefficient of restitution ($e$), post-collision relative separation speed ($v_{21cjr}$), ratio of CME2 to CME1 expansion speed ($u_{2ex}/u_{1ex}$) and a decrease in pre-collision relative approaching speed ($u_{12cjr}$). The large values of $PC_2$ primarily show a decrease in sum of expansion speed ($u_{12exs}$), and secondarily a decrease in the values of direction of impact ($\psi$) and distance ($R$) of collision site. From the bottom panel of Figure 13, we deduce that the CMEs of 2011 February 14-15 and 2012 September 25-28 showing super-elastic collision have larger values of $PC_1$, i.e., they have larger $u_{2ex}/u_{1ex}$ and $v_{21cjr}$ while smaller $u_{12cjr}$. This is also evident from the Table 3 where $u_{12cjr}$ ranges between 100 and 280 km s$^{-1}$ while its value for $e>1$ ranges only between 100 and 130 km s$^{-1}$. Further, the value of $v_{21cjr}$ for super-elastic collisions is greater than 135 km s$^{-1}$ while it is less than 45 km s$^{-1}$ for inelastic nature of collision. We may also note a weak negative correlation between $u_{12exs}$ and $e$ values. The CMEs of 2012 June 13-14 and 2011 August 3-4 have lower approaching speed as 135 km s$^{-1}$ and 145 km s$^{-1}$, respectively but they show inelastic nature of collisions which are largely away from elastic nature of collision. This is because of the fact that these CMEs also have the lower value of $u_{2ex}/u_{1ex}$ and $v_{21cjr}$ with higher $u_{12exs}$ value. The CMEs of 2012 March 4-5 have a large value of $u_{2ex}/u_{1ex}$ as 2.0 but also have a large value of $u_{12cjr}$, and therefore show perfectly inelastic nature of collision. Thus, we suggest that a super-elastic collision of the CMEs is expected with smaller $u_{12cjr}$ and simultaneously larger $u_{2ex}/u_{1ex}$ which together lead to larger $v_{21cjr}$ value.

Succinctly, among the 8 cases studied (Table 3), we find no clear dependence on direction of impact, distance of collision site, and mass ratio of different cases of the CMEs for a particular type of collision. However, some dependence of $e$ on propagation direction of CMEs is noted from the mathematical analysis described in Section 4.2. From the table, it is highlighted that the value of $e$ primarily depends, in order of priority, on $u_{12cjr}$, $u_{2ex}/u_{1ex}$, $\Delta$T, and $u_{12exs}$. Although our study is a first attempt of taking several cases of interacting CMEs, the results are limited to a large extent in the absence of enough observed cases of interacting CMEs to be analyzed for a statistically significant multi-parameter study. We also notice that the value of coefficient of restitution ($e$) is underestimated in head-on collision scenario than oblique collision scenario. Our analysis suggests that the pre-collision speeds of the centroids of CMEs (i.e., a combination of leading edge and expansion speeds) must be used in the scheme of forecasting the collision nature of CMEs.

### 4.2. *Role of the uncertainties in the observed parameters of the CMEs*

The results obtained by considering the uncertainties in the observed propagation directions, leading edge speeds and angular widths (i.e., expansion speeds) of the CMEs are noted in the Table 4, 5 and 6, respectively. We attempt to understand how the uncertainties lead to different probability for different type of collision for any selected case of interacting CMEs. The considered uncertainties in propagation directions of 2012 March and 2012 November CMEs give a likelihood of perfectly inelastic collision of around 62% and 48%, the uncertainties in their angular widths give a likelihood of perfectly inelastic collision of around 65% and 41%, and the uncertainties in their speeds give a likelihood of perfectly inelastic collision of around 98% and 100%, respectively. Similarly, the uncertainties in propagation directions of 2012 June and 2011 August CMEs give a likelihood of inelastic collision of around 65% and 77%, the uncertainties in their angular widths give a likelihood of inelastic collision of around 50% and 53%, and the uncertainties in their speeds give a likelihood of inelastic collision of around 94% and 84%, respectively. For the 2013 October

CMEs; the uncertainties in propagation directions, widths and speeds correspond to a probability of 67% for super-elastic collision, 75% for inelastic, and 50% for super-elastic, respectively. The uncertainties in the directions of 2011 February, 2012 March and 2012 September CMEs give a likelihood of super-elastic collision of around 88%, 38% and 89%, and the uncertainties in their widths give a likelihood of super-elastic collision of around 73%, 43% and 60% while the uncertainties in their speeds give a likelihood of super-elastic collision of around 89%, 61% and 49%, respectively.

From the above values for probability and values of $e$ shown in Figure 4 to 11, we note that the effect of uncertainties is so large for some cases that it is difficult to ascertain a particular nature of collision for those cases. For instance, under the uncertainties considered, three cases (CMEs of 2010 May 23-24, 2013 October 25, and 2012 September 25-28) of the CMEs show the most probable nature of collision vacillating between inelastic to super-elastic regime. Such an effect is expected for the CMEs of 2013 October 25 as its observed nature of collision was elastic, i.e. at the boundary of the inelastic and super-elastic. However, the CMEs of 2010 May show huge uncertainties in $e$ value with their directions, widths and speeds. The CMEs of 2012 September 25-28 show a large deviation from super-elastic nature due to uncertainties in pre-collision leading edge speeds of the CMEs. Such uncertainty in the $e$ value due to a change in the speed of the CMEs was not noted in Mishra et al. (2015b). We point out that in oblique collision scenario, even a reasonable uncertainty in the observed CMEs characteristics leads to a different nature of collision. The $e$ value is found to be dependent to some extent on relative propagation directions of the CMEs. However, this could not be recognized in earlier studies assuming head-on collision scenario (Mishra et al. 2014, 2015a). We envisage that a change in the propagation directions may lead to a change in the relative contribution of expansion speeds in the centroids speeds of the CMEs along the line joining their centroids. We could not make any attempt to understand if different propagation directions of CMEs causing different contact area between them have some role in deciding the collision nature. The selected cases of the CMEs which have larger variations in $e$ value are with moderate as well as lowest assigned accuracy as mentioned in Table 1. Therefore, we think that it is not solely the accuracy in the estimated pre- and post-collision kinematics which makes the value of $e$ more sensitive to the observed CMEs parameters.

From the fourth and fifth columns of Table 5, we note a larger value of the ratio of CME2 to CME1 expansion speed ($u_{2ex}/u_{1ex}$) for super-elastic collisions ($e>1$) than that for $e<1$, for all the cases of the CMEs. This is evident from second panel (from the left) of Figure 4 to 11. However, Table 6 corresponding to the uncertainties in the initial speeds of the CMEs shows no significant difference in the value of $u_{2ex}/u_{1ex}$ for super-elastic and inelastic nature of collision. The sixth and seventh columns of Table 5 show that as the ratio of summation of expansion speed ($u_{12exs}$) to relative approaching speed ($u_{12cjr}$) increases, the percentage of points (i.e., probability) having $e>1$ increases. This implies that a decrease in the approaching speed of the interacting CMEs increases the probability of their super-elastic collision. This is also evident from third panel of Figure 4 to 11. The fourth panel of Figure 4 to 11 corresponding to uncertainties in the initial speeds of CMEs leading edges, also show that a small value of relative approaching speed of centroids of the CMEs favors for an occurrence of super-elastic collision. The last column of Table 5 and 6 suggests that there remains a certain probability for inelastic collision despite a smaller approaching speed than summation of expansion speeds of the CMEs.

## 5. DISCUSSION

In earlier analysis, it has been established that the uncertainties in the mass of the CMEs has hardly any effect on the collision nature of the CMEs (Shen et al. 2012; Mishra & Srivastava 2014; Mishra et al. 2015b, 2016). This is expected from our approach as the observed post-collision speeds ($v_{1c}, v_{2c}$) of the centroids of the CMEs are modified to determine its expected values ($v_{1cth}, v_{2cth}$) to be used for estimating the value of $e$ while constraining the momentum conservation. Therefore, in the present analysis, we did not assess the effect of the uncertainties in the mass. Instead, our analysis was focused on uncertainties in other CMEs parameters. However, we do admit that the mass estimated in COR field of view may not be the actual mass at the collision sites. This is possible due to well known snowplough effect (DeForest et al. 2013; Feng et al. 2015). Further, it is difficult to know if the total masses of the interacting CMEs participate in the collision where only a part of the CME gets in contact with one another. The assumptions of a spherical structure for the CMEs and that the mass is centered at the centroid of the CMEs are idealistic, nevertheless are pragmatic for such studies.

Among 8 selected cases in our study, the SSSE method could be applied for 5 cases and thus their propagation directions could be estimated in HI field of view. We note that the directions estimates from SSSE using two viewpoint of *STEREO* is less reliable than that from the GCS model which uses an additional viewpoint from *SOHO*. Therefore, we have used the directions estimated from GCS model in COR field of view for studying the collision. This is also because during the time of occurrence of selected cases of the CMEs (except 2010 May 23-24), the separation between both *STEREO* spacecraft is either close to 180° or greater than this. For such separation of *STEREO* spacecraft, the directions estimates from SSSE has large errors and noise as described in earlier studies (Liu et al. 2010a, 2013; Mishra et al. 2014; Mishra & Srivastava 2014; Liu et al. 2016). During anti-parallel locations of *STEREO-A* and *B*, an occurrence of singularity on implementing SSSE method results in a larger uncertainty in directions even with a smaller uncertainty in the elongation measurements of the CMEs (Liu et al. 2011; Mishra & Srivastava 2013). The estimated directions from SSSE are within 10° of those derived from GCS for 3 cases of 2010 May, 2012 March and 2012 June CMEs. A larger disagreement between GCS and SSSE derived direction is around 20° for 2011 February and 2011 August cases. In our analysis, we assess the effect of ±20° uncertainties in the directions on the collision nature. Admitting the errors in tracking of the CMEs in the HI field of view and the errors from reconstruction methods (Liu et al. 2010a; Davies et al. 2013; Liu et al. 2013; Vemareddy & Mishra 2015; Mishra et al. 2015b), we believe that the observed speeds also have the uncertainties and their effect is examined in our study.

We compared the remotely derived post-collision speed values with measured speeds from in situ observations to infer the errors from remote measurements. We notice a significant disagreement between two sets of observations for the cases of 2011 February, 2012 June, 2012 September and 2013 October. In short, the in situ measured speed is smaller by 270 km s$^{-1}$ for CME1 and 140 km s$^{-1}$ for CME2 in 2012 June case, 400 km s$^{-1}$ for CME1 and 100 km s$^{-1}$ for CME2 in 2012 September case, and 150 km s$^{-1}$ for 2013 October CMEs than derived speed values from remote observations. The overestimation of speed for 2012 June, 2013 October and 2012 September CMEs from SSE or SSSE method implemented on remote observations is expected as these CMEs are receding from the observers *STEREO-A* and *B* (Liu et al. 2013; Mishra et al. 2015b). Also, the propagation direction of 2013 October CMEs suggests for their flank encounter with in situ spacecraft and this is an additional reason for the discrepancy between remote and in situ measured speed values for this case. The in situ measured speed of 2011 February CMEs is 100 km s$^{-1}$ larger than its post-collision speeds measured from remote observations. This is possible if the CMEs over-expand before reaching to L1. The remotely measured kinematics and its connection with in situ measured values and the disagreement therein for the CMEs of 2011 February 14-15, 2010 May 23-24, 2012 November 9-10, and 2012 September 25-28 are discussed in earlier studies of Mishra & Srivastava (2014); Lugaz et al. (2012); Mishra et al. (2015a) and Mishra et al. (2015b), respectively.

The large discrepancy between remote and in situ measured post-collision speeds may reduce the accuracy of our analysis for collision nature for the cases (e.g., 2012 September 25-28 and 2012 June 13-14 CMEs) where CMEs have little or no chance of making their flank encounter with in situ spacecraft and collision takes place largely away from the Sun. The appraisal of errors in our analysis would be difficult as the errors in pre-collision speeds of the CMEs cannot be inferred from comparison of remote and in situ measurements. If there is equal proportion of errors in the pre- and post-collision speeds then they get nullified by one another in the calculation of the value of coefficient of restitution. We note that the implemented SSSE reconstruction method on HI observations of 2012 September CMEs have overestimated the remotely measured post-collision speed of CME1 to a greater extent than CME2. This implies that it is likely that not as much momentum was transferred to CME1 as is found from the kinematics profiles and taken for the analysis. It is therefore very likely that this event was not as super-elastic (or not at all) as what was found in our analysis taking oblique collision scenario. This is in agreement to results listed in Table 6 where this event shows likelihood of 50% for inelastic nature of collision under the uncertainties of $\pm 100$ km s$^{-1}$ in the observed post-collision speed. Similarly for the case of 2012 June CMEs, the SSSE method over-estimated the post-collision speed of CME1 to a larger extent than that of CME2. Therefore, the nature of collision of this case could have shifted further towards close to perfectly inelastic. Therefore, the approach should be to ascertain only a finite probability for a particular nature of collision. We also keep in mind that only based on good agreement between two set of remote and in situ observations, the possibility of errors in the speeds derived using remote observations cannot be excluded. This is because of difficulty in accurately quantifying the possible deceleration, acceleration, deflection, over-expansion of CMEs beyond the collision site and trajectory of in situ spacecraft through the CME. Thus it is possible that complex interactions during the Sun-to-Earth propagation may not be revealed by in situ measurements alone. Further, the inconsistency in speed values from remote and in situ measurements may be partially due to the fact that remotely tracked feature of a CME has not been intercepted by the in situ spacecraft.

For all the selected cases, we noticed a deceleration of the leading edge of the following CME well before an acceleration of the leading edge of the preceding CME. This is partly because the preceding CME may act as a magnetic obstacle for the following CME (Temmer et al. 2012; Mishra & Srivastava 2014), and thus a remote interaction between them starts before the actual collision takes place. This is also partly because a thrust from the following CME on the rear part of the preceding CME requires some time (i.e. minimum Alfvenic crossing time for the CME) to arrive at its leading edge. We also point out that the identification of collision phase based on the observed exchange in kinematics of the CMEs leading edges may involve errors and creates the uncertainties in our analysis. The marked start of collision in our study is postponed than the actual contact between the CMEs as described in Mishra et al. (2016). This leads to an over-estimation of $e$ value to some extent which may be nullified by an underestimation of $e$ value caused by ignoring the contribution of CME2 driven shock to accelerate CME1 in our study. Such errors causing the competitive effects of over- and under-estimation of the value of $e$ need to be explored further. It is expected that a collision phase is complex involving different time scales for compression by shock, subsequent expansion, exchange of momentum and magnetic reconnection. Therefore, we do realize that large-scale magnetically structured plasmoids would not collide as ordinary objects.

Some separation between leading edge of the preceding and the following CME at the beginning of the collision is expected due to finite size of the preceding one. Carefully inspecting the estimated distance profiles of leading edge of the CMEs for 4 cases (i.e., 2011 February, 2012 June, 2013 October and 2011 August), we note that the separation is within 5 $R_\odot$. However, this separation is ranging from 20 $R_\odot$ to 25 $R_\odot$ for other 4 cases (i.e., 2010 May, 2012 March, 2012 November and 2012 September). The obtained separation between leading edge of CME2 and CME1 implies much smaller diameter of CME1 than reported for CMEs in earlier studies (Leitner et al. 2007; Gulisano et al. 2010; Wang et al. 2015). It is understood that the separation between the leading edge of CME2 and CME1 depends not only on the size of CME1 but also on their propagation directions. Thus, the direct inference of the size of preceding CME only from the separation between leading edges of colliding CMEs may be inaccurate, in the case of oblique collision where projection effect may be significant. Further, it is difficult to ensure that tracked features of the CMEs in the J-maps correspond to the outermost portion of the CMEs. Also, the possible compression of the colliding CMEs may prevent us measuring the exact diameter of CME1. Taking all these points into account, the inferred smaller size of the preceding CME is not quite surprising. However, on admitting the possibility of errors in marking the collision phase, the inferred smaller size of CME1 may imply that beginning of collision should be marked little earlier. Due to this fact the value of $e$ would be overestimated to some extent in our analysis where the collision phase is marked noticing the speed variations of the CMEs than separation between their measured leading edges.

CME-CME interaction is actually a three-dimensional phenomenon. Although the kinematics of the CMEs is estimated using SSE or SSSE methods of 3D reconstruction, the J-maps are made along the ecliptic only. This means that the kinematics used even in the oblique collision scenario represents the collision of only a part of the CMEs as shown by Temmer et al. (2014) that colliding CMEs have different speeds at different position angles. In our study for the cases where SSE method is used, either post-collision directions or deflection of the CMEs during collision is not used for estimating their observed speeds. However, while solving the equations for the collision, we determined the post-collision directions and modified the observed post-collision speeds to be taken for collision analysis. The theoretical estimation of expected post-collision speeds for the analysis suggests that the deflection of the CMEs cannot be completely overlooked. Since, the elongation measurements for an observer and the observed propagation directions of the CMEs are linked, the true effect of a change between pre- and post-collision directions on the speed is difficult to assess. Further, we have not considered the possible rotation and deflection of the CMEs and focus on the linear speeds of the centroids. Also, we have ignored the contribution of solar wind in acceleration or deceleration of the CMEs during the collision phase (Shen et al. 2012, 2013). Such assumptions probably lead to significant errors for the cases of the CMEs like Feb 14-15 and Sep 25-28 which have significantly longer collision phase. Neglecting the effect of solar wind probably induces errors for 2011 August 3-4 CMEs where their post-collision speeds are derived from the in situ observations while the collision occurred around 145 $R_\odot$ from the Sun.

From our analysis of observed cases of the CMEs, it is clear that a significantly larger expansion speed of the following CME than preceding one contributes in reducing the relative approaching speed of centroids of the CMEs. The larger expansion speed of the following CME implies a larger internal pressure inside (Wang et al. 2009) and probably refers to harden the CMEs. It has been found in experiments that a collision of hard ceramic spheres with softer polycarbonate plates (Louge & Adams 2002; Kuninaka & Hayakawa 2004) is super-elastic in nature. It seems that some of the magnetic and thermal energy of the following CMEs gets converted into macroscopic kinetic energy of the CMEs to make the collision super-elastic. We are inclined to propose that the internal pressure of the CMEs indirectly displays the physical nature of the macroscopic expanding plasma blobs. Therefore, the different physical characteristics of the CMEs plasma may lead to different types of collision. The fair dependence of nature of collision on the duration of collision obtained in our study possibly indicates for the role of plasma process in CME-CME collision.

The long duration of collision intend to show super-elastic collision where additional kinetic energy would have been produced through magnetic reconnection. A through understanding of the physics and plasma processes responsible for super-elastic collision may involve the role of magnetic pressure and orientation of flux ropes of the CMEs (Lugaz et al. 2013). Despite doing an extensive data analysis for a total of 8 cases, we could not establish a sufficient condition for super-elastic collision to occur. Several limitations of such a study using the imaging observations are discussed in earlier studies (Shen et al. 2012; Lugaz et al. 2013; Mishra et al. 2016).

To address some of the limitations related to CMEs driven shock (Vandas et al. 1997; Lugaz et al. 2005), heating, compression (Shen et al. 2013, 2016), role of reconnection and over-expansion (Lugaz et al. 2013), the numerical simulation of such CMEs may be helpful. Understanding the role of oblique collision of CMEs in deciding their deflection and elasticity using simulation studies (Schmidt & Cargill 2004; Xiong et al. 2009) is also required. It is possible to get some clues on physical processes during the collision if there would be some in situ measurements of CMEs plasma shortly before and after the collision site. The present study emphasizes to state only the probable nature of a collision leaving room for the uncertainties. An analysis for several cases by combining the simulations and observations for the same colliding CMEs are important for making progress in understanding the nature of collision of CMEs as well as space weather studies.

## 6. CONCLUSION

Our study emphasizes the possibility of a large uncertainty in the calculated value of coefficient of restitution ($e$) from the observed CMEs characteristics. Such an uncertainty is not obviously noticed while considering the head-on collision scenario where the value of $e$ is often underestimated. This is evident as on taking into account the uncertainties for oblique collision scenario, the nature of collision of 3 cases (i.e., 2010 May, 2013 October and 2012 September CMEs) among the 8 cases of the CMEs, could not be ascertained decisively. We suggest that nature of collision of the CMEs should only be determined with a finite probability for a specific nature. We note that direction of impact, distance of a collision site from the Sun, and mass ratio of the CMEs do not favor for a particular nature of collision. The decrease in the pre-collision approaching speed of CMEs along the line joining their centroids with an increase in ratio of CME2 to CME1 pre-collision expansion speed, give a large probability for super-elastic collision. The study concludes that the large expansion speed of the following CME than that of the preceding CME, giving a relatively lower approaching speed before the collision and a higher separation speed after the collision, tends to increase the probability for super-elastic collision (Shen et al. 2012, 2016). Thus the expansion speed of the CMEs plays a greater role than any other CMEs parameters. Our study shows the dependence of the calculated nature of collision on propagation directions, angular sizes and leading edge speeds of the CMEs. This is probably because these parameters of the CMEs indirectly alter the relative contributions of expansion speeds in the leading edge speeds of the CMEs and thus their relative approaching speeds of the centroids. However, these uncertainties in the CMEs parameters do not alter the physical nature of the collision for any selected cases. The physical processes responsible for probably converting the magnetic energy into kinetic energy of the CMEs, to make a collision super-elastic, need to be addressed in details.

We acknowledge the UK Solar System Data Center for providing the processed Level-2 *STEREO*/HI data. The work is supported by the NSFC grant Nos. 41131065, 41574165, and 41421063. W. M. is supported by the Chinese Academy of Sciences (CAS) President's International Fellowship Initiative (PIFI) grant No.

2015PE015. We thank the referee for his/her insightful comments. W.M. also thank A. K. Awasthi and A. Raghav for helpful discussions.

## APPENDIX-A

Our approach considers that two successively launched CMEs (CME1 and CME2) having different angular half-width ($\omega_1$ and $\omega_2$) are propagating as expanding bubbles in two different directions ($\phi_1$ and $\phi_2$) from the Sun-Earth line and along the propagating directions $\alpha_1$ and $\alpha_2$ relative to the line joining their centroids at the instant of collision.

Thus, we will get

$$\cos(|\phi_1 - \phi_2|)\sin(\alpha_1) + \sin(|\phi_1 - \phi_2|)\cos(\alpha_1) = \frac{\sin(|\phi_1 - \phi_2|) - \sin(\omega_2)\sin(\alpha_1)}{\sin(\omega_1)} \quad (1)$$

$$\alpha_2 = \alpha_1 + |\phi_1 - \phi_2|$$

Under the usual notations and conventions used in this paper, for CME1; $u_{1c} = u_1 - u_{1ex}$ and $u_{1ex} = u_1 \sin(\omega_1)/[1 + \sin(\omega_1)]$. Similarly, it will be for CME2. The post-collision directions of propagation of the CMEs relative to Sun-Earth line is $\phi'_1$ and $\phi'_2$ and they are $\beta 1$ and $\beta 2$ relative to the line joining their centroids. With theoretically determined expected post-collision speeds of the centroids ($v_{1cth}, v_{2cth}$) of the CMEs using a certain value for $e$ which together allow the momentum to be conserved for the collision, we will get Equations 2, 3 and 4.

$$v_{1cth}\cos(\beta_1) = \frac{m_1 u_{1c}\cos(\alpha_1) + m_2 u_{2c}\cos(\alpha_2) - m_2 e[u_{1c}\cos(\alpha_1) - u_{2c}\cos(\alpha_2)]}{m_1 + m_2}$$
$$v_{2cth}\cos(\beta_2) = \frac{m_1 u_{1c}\cos(\alpha_1) + m_2 u_{2c}\cos(\alpha_2) + m_1 e[u_{1c}\cos(\alpha_1) - u_{2c}\cos(\alpha_2)]}{m_1 + m_2} \quad (2)$$

$$u_{1c}\sin(\alpha_1) = v_{1cth}\sin(\beta_1)$$
$$u_{2c}\sin(\alpha_2) = v_{2cth}\sin(\beta_2) \quad (3)$$

$$\phi' = \phi + (\beta - \alpha) \text{ for } \phi_1 < \phi_2$$
$$\phi' = \phi - (\beta - \alpha) \text{ for } \phi_1 > \phi_2 \quad (4)$$

For the oblique collision scenario considered, the speed of the centroid of CME1 along the line joining the centroids of both the CMEs will be $u_{1cj} = u_{1c}\cos(\alpha_1)$ and similarly for CME2. For collision to occur, $[u_{2cj} + u_{2ex}] \geq [u_{1cj} - u_{1ex}]$ and $|(\phi_1 - \phi_2)| \leq (\omega_1 + \omega_2)$ must be satisfied. Using the above Equations, the value of $v_{1cth}$ and $v_{2cth}$) for a definite value of the coefficient of restitution ($e$) is determined. Thereafter the expected values of post-collision speeds of leading edge of the CMEs ($v_{1th}, v_{2th}$) is determined which is compared with leading edge speeds ($v_1, v_2$) as observed. The best suited value of $e$ is attributed to the nature of collision of the selected CMEs for which the deviation (i.e. $\sigma = \sqrt{[(v_{1th} - v_1)^2 + (v_{2th} - v_2)^2]/2}$) between the observed and the expected post-collision leading edge speeds is minimum. The pre-collision relative approaching speed and post-collision relative separation speed of the centroids of CME1 and CME2 is defined as $u_{12cjr} = |u_{2cj} - u_{1cj}|$ and $v_{21cjr} = |v_{1cj} - v_{2cj}|$, respectively.

## APPENDIX-B

To visualize the pattern of similarity/difference between the observations points and variables in Table 3, we used the most popular multivariate statistical method called as principal component analysis (PCA) (Pearson 1901; Hotelling 1933; Jolliffe 2002). The method reduces the dimensionality of a data set while preserving as much 'variability' (i.e. statistical information) as possible. The important information from the data set is expressed as a set of new uncorrelated orthogonal variables called principal components (PCs). Finding such new variables (PCs) reduces to solving an eigenvalue and eigenvector problem or singular value decomposition of the data matrix. The first principal component (i.e. PC$_1$) is the axis that spans and captures the direction of the most variation in the data. The PC$_2$ is the axis that spans the second most variation in the data, and similarly there are principal components for each variables (i.e., dimension) in the original data. Following the terminology of Abdi & Williams (2010), the value of new variables (PCs) for the observations are called 'factor scores'. These 'factor scores' are seen geometrically as the 'projections' of the observations onto the principal components. Thus, a projection matrix as having the eigenvectors of the data matrix in each row is used for 'loadings' of the original variables on the PCs.

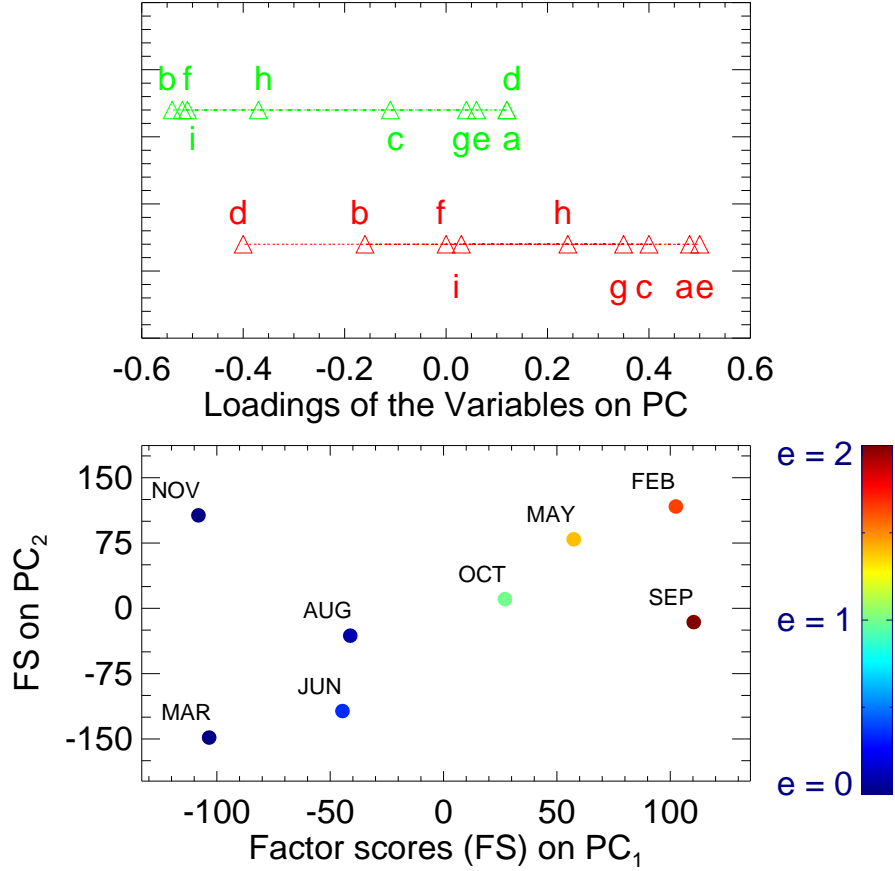

**Figure 13.** The output from principal component analysis (PCA) is shown. From top to bottom, the first panel shows the loadings of the variables on two principal components PC$_1$ with red and PC$_2$ with green. The second panel shows the variations of factor scores of the observations (for CMEs cases of 2011 February, 2012 June, 2010 May, 2012 March, 2012 November, 2013 October, 2011 August and 2012 September as listed in Table 1) on PC$_1$ and PC$_2$.

From table 3, we take 9 variables as $e$, $u_{12exs}$, $u_{2ex}/u_{1ex}$, $u_{12cjr}$, $v_{21cjr}$, $\psi$, $\Delta T$, $m_2/m_1$, and $R$ and denote them with $a, b, c, d, e, f, g, h$, and $i$, respectively, for all the 8 observations of the selected CMEs events. This data set made a data matrix of 8 rows and 9 columns. The 9 eigenvalues for this data set are as 3.96, 2.67, 1.32, 0.49, 0.32, 0.20, 0.02, 0, 0 corresponding to 9 PCs associated with them. We note that PC$_1$ and PC$_2$, having eigenvalues greater than 2, account for 44% and 29.7% variance in the data, respectively. The other 7 components (i.e., PC$_3$ to PC$_9$) taken together account for only 25% variations in the data. Therefore we may keep only the first two PCs for further consideration. The eigenvectors corresponding to two eigenvalues represent the loading of the variables on the PC$_1$ and PC$_2$. The loading of all the 9 variables, i.e., from $a$ to $i$ on PC$_1$ is [0.48, -0.16, 0.40, -0.40, 0.50, 0, 0.35, 0.24, 0.03] and shown with red in the top panel of the Figure 13. The higher the loadings, the more important that variable is to the component. This shows that PC$_1$ contrasts the variable $d$ with the variables $a$, $c$, and $e$, as well as it captures the variations in these variables. Similarly the loading of variables on PC$_2$ are as [0.12, -0.54, -0.11, 0.12, 0.06, -0.52, 0.04, -0.37, -0.51] and shown with green in the top panel of Figure 13. PC$_2$ captures the variations in the variables $b$, $f$ and $i$. Using the loading of the variables, we determined the factor scores of all the 8 observations (i.e., CMEs events from Feb 14-15 to Sep 25-28 as noted in Table 1) on PC$_1$ and PC$_2$. The obtained factors scores on these first two PCs are displayed in bottom panel of Figure 13. The result derived from this analysis is emphasized in Section 4.1.